\documentclass[
    aps,
    prapplied,
    reprint,
    groupedaddress
]{revtex4-2}

\usepackage{hyperref}
\usepackage{blkarray}
\usepackage{comment}
\usepackage{dcolumn}
\usepackage{booktabs}
\usepackage{amsmath,amssymb}
\usepackage{graphicx}
\usepackage{soul,color}
\usepackage[table,xcdraw]{xcolor}  
\usepackage{algorithmic}
\usepackage{algorithm}
\usepackage{array}
\usepackage{multirow}
\usepackage[normalem]{ulem}
\usepackage{colortbl}
\usepackage{makecell}

\definecolor{Gray}{gray}{0.5}

\hypersetup{
    colorlinks = true,
    linkcolor = blue,
    citecolor = blue,
    filecolor = blue,
    urlcolor = blue
}

\begin{document}

\preprint{APS/123-QED}

\title{Adaptive Variation-Resilient Random Number Generator for Embedded Encryption}
\author{Furqan Zahoor$^{1\dagger}$, Ibrahim A. Albulushi$^{2\dagger}$, Saleh Bunaiyan$^{2,3}$, Anupam Chattopadhyay$^4$, Hesham ElSawy$^5$, and Feras Al-Dirini$^{5,6}$}\email{Corresponding Author: aldirini@mit.edu}

\affiliation{$^1$Department of Computer Engineering, King Faisal University, Al-Ahsa, Saudi Arabia}

\affiliation{$^2$Electrical Engineering Department, King Fahd University of Petroleum $\&$ Minerals,  Dhahran, Saudi Arabia}

\affiliation{$^3$Electrical and Computer Engineering, University of California, Santa Barbara, Santa Barbara, CA, USA}

\affiliation{$^4$College of Computing $\&$ Data Science, Nanyang Technological University, Singapore, Singapore}

\affiliation{$^5$School of Computing, Queen's University, Kingston, ON, Canada}

\affiliation{$^6$Research Laboratory of Electronics, Massachusetts Institute of Technology, Cambridge, MA, USA}

\affiliation{$^\dagger$These authors contributed equally to this work}

\date{\today}

\begin{abstract}
With a growing interest in securing user data within the internet-of-things (IoT), embedded encryption has become of paramount importance, requiring light-weight high-quality Random Number Generators (RNGs). Emerging stochastic device technologies produce random numbers from stochastic physical processes at high quality, however, their generated random number streams are adversely affected by process and supply voltage variations, which can lead to bias in the generated streams. In this work, we present an adaptive variation-resilient RNG capable of extracting unbiased encryption-grade random number streams from physically driven entropy sources, for embedded cryptography applications. The system's key feature is its adaptive digitizer with an adaptive reference voltage. As a proof of concept, we employ a stochastic magnetic tunnel junction (sMTJ) device as an entropy source. The impact of variations in the sMTJ is mitigated by the adaptive digitizer, which generates an adaptive short-term average reference voltage that dynamically tracks any stochastic signal drift or deviation, leading to unbiased random bit stream generation. The generated bit streams, due to their higher entropy, then only need to undergo simplified post-processing. A prototype of the adaptive RNG system was experimentally implemented using discrete electronic components and an FPGA for entropy source emulation. Statistical randomness tests based on the National Institute of Standards and Technology (NIST) test suite are conducted on bit streams obtained using the simulations as well as the discrete electronic component implementation, demonstrating that the bit streams consistently pass all 16 tests of the NIST SP 800-22 test suite with a 100\% pass rate. Leveraging its simplified post-processing, the adaptive RNG shows consistent operation across a wide range of throughputs from 5 to 182 Mbps. Our results validate the system's ability to generate encryption-grade random bit streams at reduced hardware cost for IoT devices. 

\end{abstract}

\maketitle

\section{Introduction}
\label{Introduction}

Demand for hardware security is growing rapidly due to the unprecedented growth of the Internet of Things (IoT) \cite{raj2024puf}. Data encryption is becoming more important, with a drive to move it towards the device side and away from the cloud. This move provides higher levels of security; however, it imposes stringent size and energy constraints on the hardware. These constraints are also coupled with stringent requirements for encryption quality, imposing a challenging design trade-off between hardware compactness and encryption quality \cite{clemente2024lightweight}.

\begin{figure}[!t]
    \centering
    \includegraphics[width=0.95\columnwidth]{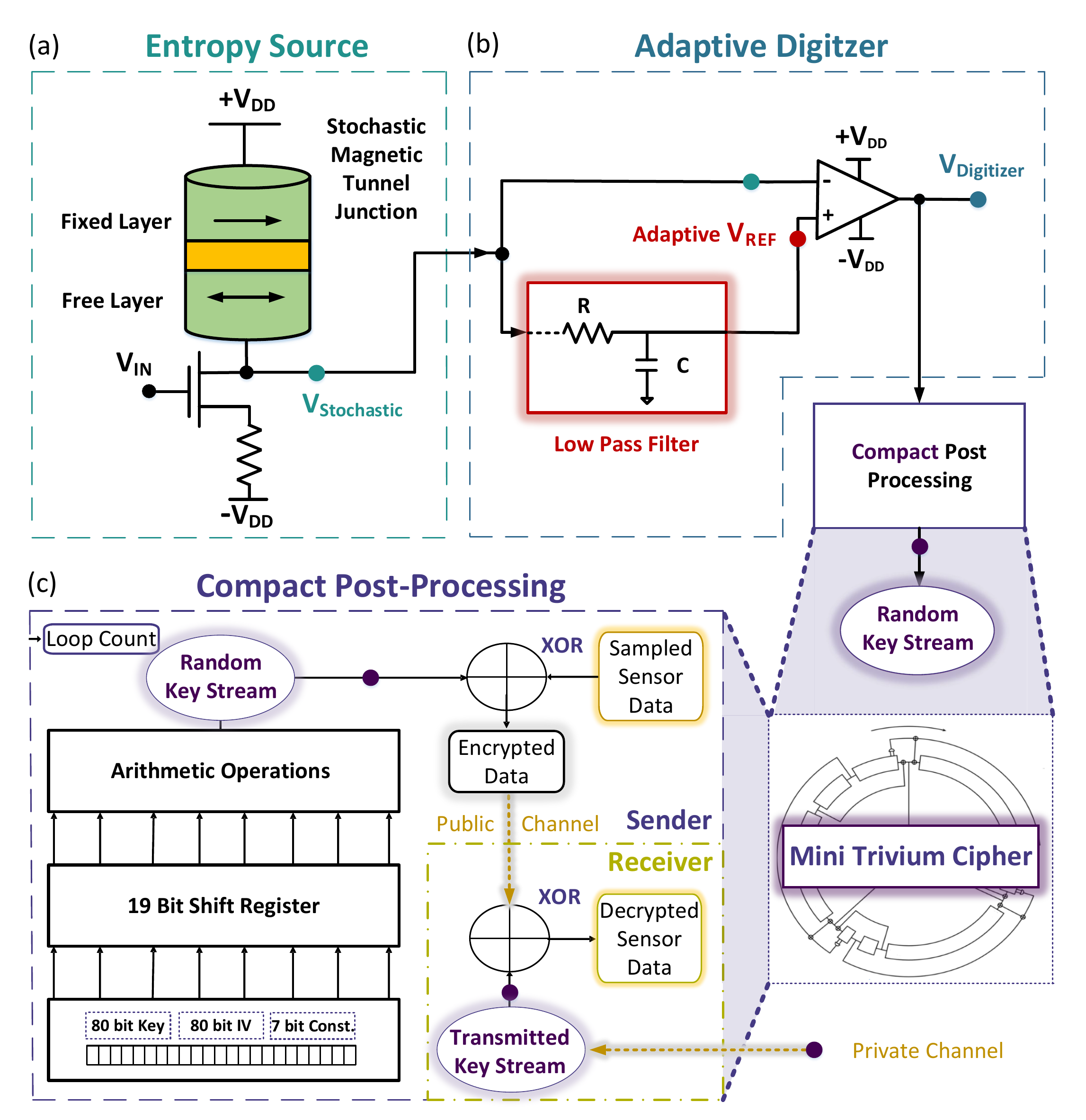}
    \caption{The adaptive RNG comprising (a) an entropy source, (b) an adaptive digitizer with a low-pass-filter generated moving reference voltage, and (c) compact post processing. }
    \label{Fig 1}
\end{figure}

For cryptographic applications, the encryption key is the main secret necessary for data decryption, since it is assumed that adversaries are aware of the encryption algorithm used \cite{de2017variation,onizawa2020,rajendran2024harnessing}. Therefore, it is crucial to generate keys through random processes that make them unlikely to be guessed. Such procedures are implemented using subsystems known as Random Number Generators (RNGs), and are of two categories: True-RNGs (TRNGs) and Pseudo-RNGs (PRNGs). The latter include circuits that implement mathematical or computational algorithms that are typically used to generate random sequences; however, such sequences are deterministic and can be regenerated if the initial `seed' is known, making them susceptible to prediction penetration. On the other hand, TRNGs are circuits that produce random bits based on physically random phenomena in emerging device technologies, making them more reliable for applications where security is of great concern \cite{frustaci2024high,singh2023hardware}, such as wireless communication \cite{liu2016low, zahoor2024novel} and remote sensing \cite{WirelessGeophoneAccess, bunaiyan2021real, bunaiyan2022neuro, P-Sensing-EDTM, PatentPSensing, P-SensingISCAS}. 

Hardware TRNG implementations are based on utilization of devices with unique properties or exploitation of intrinsic noise in classical electronic devices. TRNGs have become essential for cryptography-related IoT applications \cite{PatentAdaptiveTRNG}, as well as for probabilistic computing \cite{camsari2017,bunaiyan2022mtj,PatentTunablePBit,PatentDecoupledPBit,PatentPNeurons,P-NeuronsISCAS}. The challenge in using TRNGs is that they are adversely affected by variations in the fabrication process of the device technology used \cite{qu2018variation, SNWMemristor, RSQMemristorScaling}, leading to bias in generated bit streams. Moreover, in IoT devices, which are mainly battery powered, fluctuations in the supply voltage can also adversely affect TRNGs \cite{wallace2016toward,baturone2022unified}.

In this work, we present an adaptive RNG that extracts high-quality unbiased random bit streams from a time-varying stochastic signal obtained from an entropy source. The system is primarily designed for bit steam extraction from a true-random entropy source, such as a stochastic Magnetic Tunnel Junction (sMTJ), shown in Fig. \ref{Fig 1} (a), but we will also show that it can still perform satisfactorily well with a pseudo-random source, such a Linear Feedback Shift Register (LFSR). The complete block diagram of the proposed system is shown in Fig. \ref{Fig 1}. In our proposed adaptive RNG, an adaptive digitizer is employed to eliminate any bias, enhancing the resilience of the system to process-induced and supply-voltage variations in true-random entropy sources, and enhancing the randomness qualities of pseudo-random entropy sources. 

\begin{figure*}
 \centering
    \includegraphics[width=7 in]{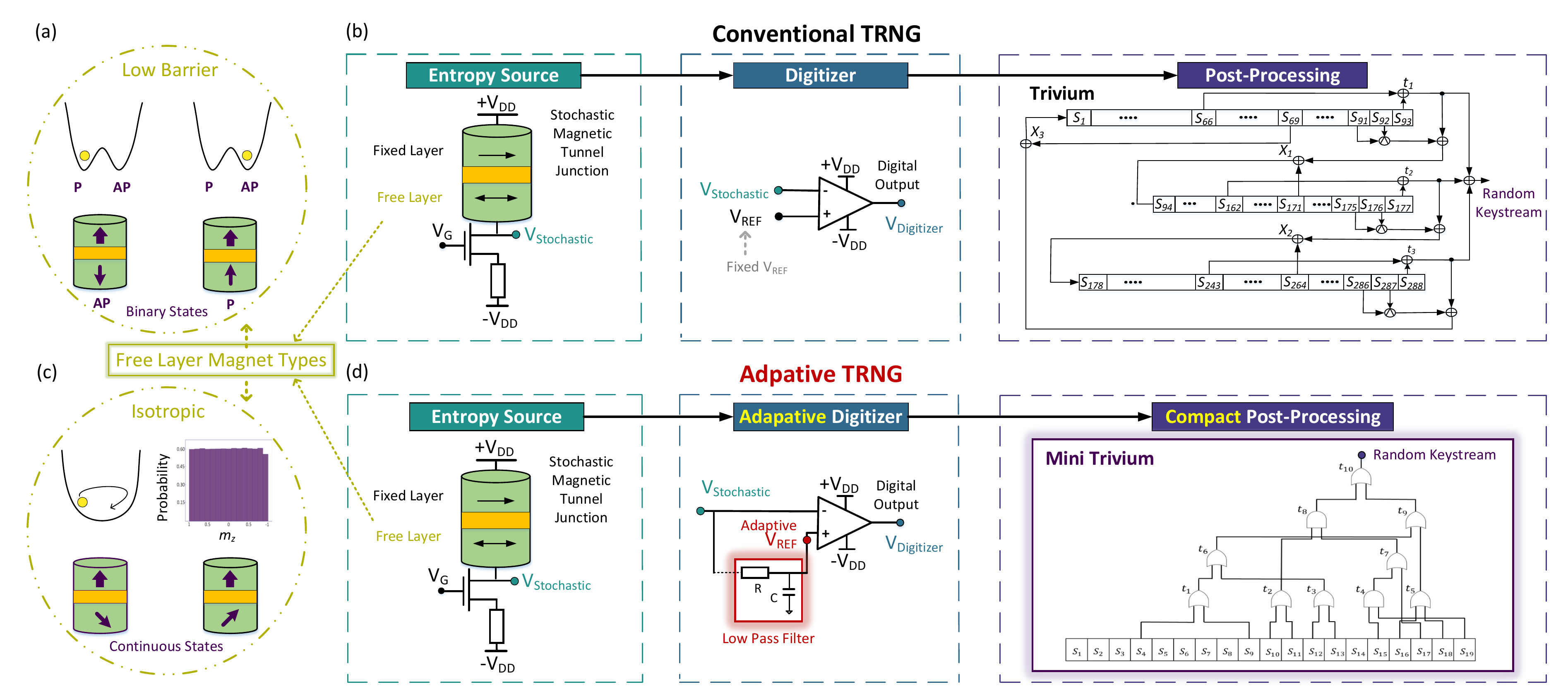}
    \caption{Comparison between conventional TRNGs and the proposed adaptive TRNG. Illustration of magnetization dynamics in stochastic MTJs for (a) low-barrier magnets and (c) isotropic magnets. The uniform probability distribution of the magnetization in the vertical z-direction ($m_{z}$(t)) depicts characteristics of isotropic sMTJs (inset of (c)). Schematic block diagram Illustration of (b) a conventional TRNG system and (d) the proposed adaptive TRNG system with the adaptive digitizer and the Mini Trivium compact post-processing blocks.}
    \label{Fig 2}
\end{figure*}

Due to the enhanced randomness qualities of the generated bitstreams, we show that light-weight post-processing is sufficient for acceptable encryption-grade randomness, allowing further savings in  hardware. This is demonstrated using both a conventional Trivium cipher \cite{de2008, tian2009, montoya2018}, shown in Fig. \ref{Fig 2} (b), and a simplified more compact version of it, the Mini-Trivium, shown in Fig. \ref{Fig 2} (d). Schematic illustrations of a conventional and the proposed adaptive RNG are depicted in Fig. \ref{Fig 2}. 

Section II discusses entropy sources, Sec. III presents an adaptive TRNG with a true-random sMTJ entropy source and Sec. IV presents an adaptive RNG with a pseudo-random LFSR entropy source with experimental implementation and compares the presented adaptive RNG with state-of-the-art RNG designs. Finally, Sec. V concludes the paper.

\section{The Entropy Source}

Entropy sources are a core component in RNGs, as they provide the source of randomness. Deterministic PRNGs employ pseudo-random digital sources in the form of shift registers and are regarded as RNGs with low-quality randomness. On the other hand, TRNGs based on conventional complementary-metal-oxide-semiconductor (CMOS) technology commonly employ noise in devices, such as thermal noise \cite{tokunaga2008true} or random telegraph noise (RTN) \cite{brederlow2006low}, as their physical entropy source and generally require oscillator circuits. Moreover, CMOS-based TRNGs need intricate extraction and complex post-processing circuits to mitigate output correlation and bias. Additionally, they are also highly affected by variations in process, voltage, and temperature (PVT) \cite{srinivasan20102}, necessitating entropy-tracking feedback loops \cite{gong2019true}.

Emerging devices utilizing stochastic switching mechanisms, like magnetic switching \cite{dubovskiy2024one}, resistive switching \cite{akbari2023true}, and phase change \cite{piccinini2017self}, have been proposed as promising entropy sources for TRNGs due to their inherent stochastic nature. Notably, devices with stochastic temporal dynamics driven by fluctuations in thermal energy, such as stochastic magnetic tunnel junctions (sMTJs) \cite{borders2019, PhysRevApplied.17.014016}, do not require voltage pulsing or forced switching mechanisms to trigger their stochastic response, making them excellent entropy sources for TRNGs. However, as stated earlier, a major challenge in these device technologies is their limited resilience to process-induced and supply voltage variations.

In order to demonstrate the utility of our adaptive RNG, we test it with an sMTJ as a true-random source and an LFSR as a pseudo-random source. The first is investigated in Sec. III and the latter is implemented and validated experimentally in Sec. IV.

\subsection{Stochastic MTJ Model}
\label{Stochastic MTJ Model}
Prior to presenting our design for an sMTJ-based adaptive TRNG, in order to be able to study the system's behavior when employing an sMTJ entropy source, a model for the sMTJ is needed. This is achieved by a two-fold process. First, the sMTJ resistance fluctuations are simulated using MATLAB, and the resulting sMTJ behavior is fed into SPICE as a variable resistor with time-varying conductance. 

The dynamics of the stochastic switching of the sMTJ are controlled by the fluctuations of the magnetization state ($m_{z}$(t)) of the free magnet layer, that fluctuates from 1 to -1 continuously, where 1 refers to a parallel state in reference to the fixed magnet layer, while -1 is an antiparallel state, as shown in Fig. \ref{Fig 2} (a). These fluctuations give rise to an sMTJ whose time-varying conductance ($G(t)$) is defined as:

\begin{equation}
G(t) =  G_{0} \bigg[ 1 + m_{z} (t)  \frac{TMR}{2 + TMR} \bigg]
 \label{eq:HD}
\end{equation}
where $G_{0}$ is the average conductance, and $TMR$ is the tunneling magnetoresistance ratio. The $TMR$ depends on the ratio between the parallel ($R_{P}$) and the anti-parallel ($R_{AP}$) resistance states of the sMTJ and is defined as:

\begin{equation}
TMR =  \frac{R_{AP}- R_{P}}{R_{P}}
\label{eq:TMR}
\end{equation}
 
MATLAB is used to emulate the sMTJ’s dynamic magnetization state $m_{z}$(t) and its probability distribution. The resultant $m_{z}$(t) vector is then used in solving eq. (1) to obtain the time-varying vector $G(t)$, which is received by SPICE software as a time-varying resistance according to $G(t)$.

Fig. \ref{Fig 2} (a) illustrates the switching mechanism of the sMTJ device employing a low barrier free magnet layer, having two stable magnetic configurations: parallel (P) and antiparallel (AP) states. In our analysis, we have employed an isotropic magnet for the free layer, owing to its uniform $m_{z}$(t) distribution as depicted in Fig. \ref{Fig 2} (c), which makes it a favorable entropy source due to the higher degrees of freedom offered by the isotropic free layer magnet \cite{hassan2021quantitative}.

\section{Adaptive TRNG with a True-Random Entropy Source}
This section presents an example TRNG design based on the adaptive RNG concept, by employing the s-MTJ true-random entropy source discussed in Sec. II. After the entropy source, the two remaining components of the adaptive TRNG system are: (1) the adaptive digitizer and (2) the post-processing circuit. These are presented in subsections A and B, respectively, followed by simulation results and variation-resilience analysis in subsections C and D, respectively.

\subsection{Adaptive Digitizer}
\begin{figure} [t]
   \centering
  \includegraphics[width=\columnwidth]{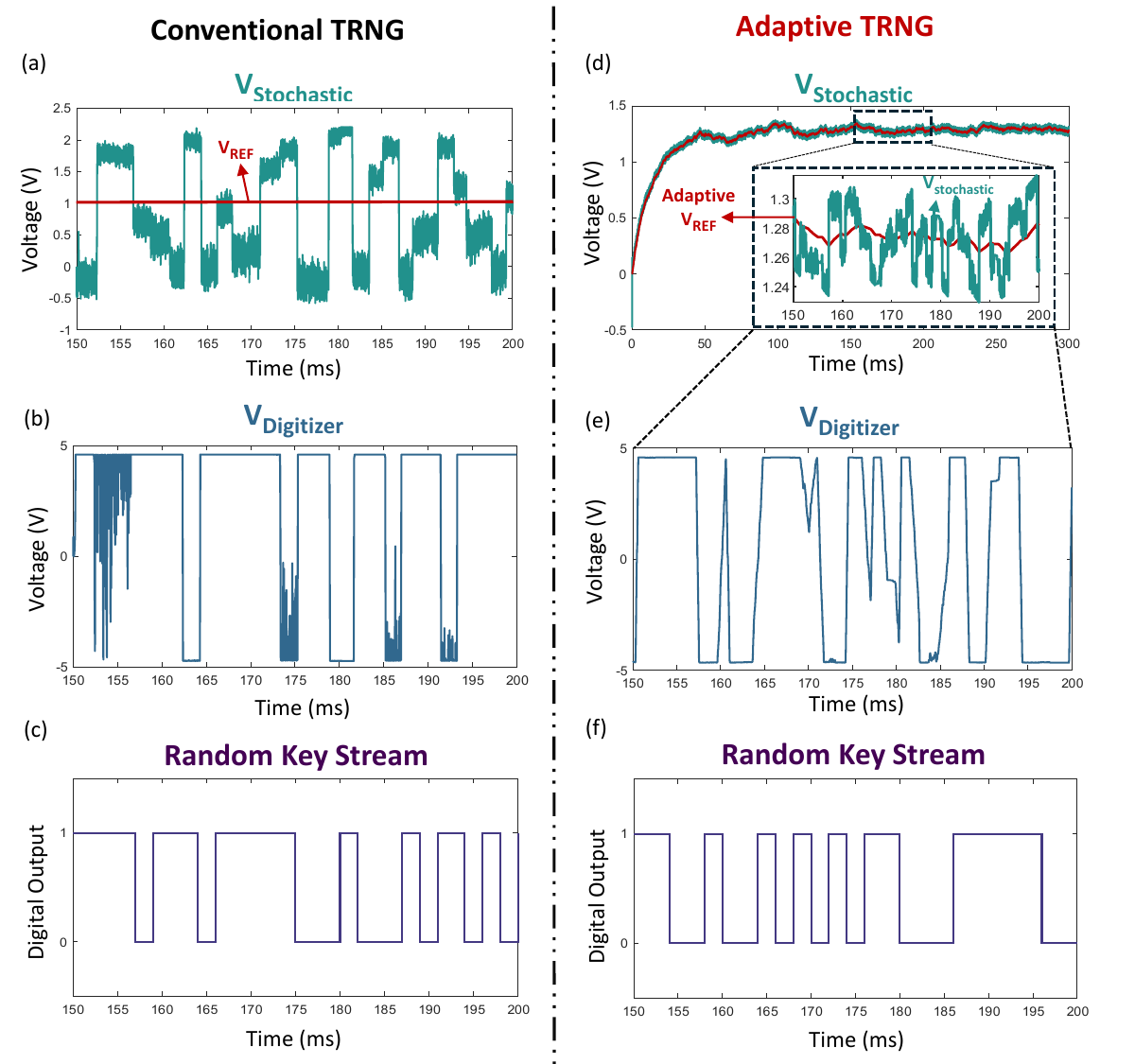}
  \caption{Simulation results for an sMTJ-based adaptive TRNG. Plots of (a) $V_{Stochastic}$ vs. time, (b) $V_{Digitizer}$ vs. time, and (c) the generated random key stream vs. time, for the conventional TRNG. Plots of (d) $V_{Stochastic}$ vs. time, (e) $V_{Digitizer}$ vs. time and (f) the generated random key stream vs. time, for the adaptive TRNG.}
  \label{Fig 3}
\end{figure}

Once a stochastic signal is generated by the entropy source, the signal needs to be converted from analog to digital, and this is the role of the digitizer. This is usually done by comparing the analog stochastic voltage signal with a reference analog voltage ($V_{REF}$) using a comparator that converts voltages above $V_{REF}$ to high output voltages (ideally $V_{DD}$: a digital 1) and voltages below $V_{REF}$ to low output voltages (ideally GND: a digital 0). Ideally, the desired output of the digitizer would be a random bit stream that is 50 \% ones (1's) and 50 \% zeros (0's) on average. Based on this, $V_{REF}$ is a critical design parameter. In our design, instead of applying a fixed $V_{REF}$ to the comparator, we design a low-pass filter (LPF) circuit that generates an adaptive moving $V_{REF}$ from the stochastic signal ($V_{stochastic}$) corresponding to its short-term average, as shown in Fig. \ref{Fig 3} (d). 

An important novelty of our system lies in this $V_{REF}$ that adapts to changes in the stochastic signal, such as voltage drift or mismatch from the designed-for voltage levels due to process-induced or supply voltage variations, and represents a real-time short-term average of the stochastic signal itself. This is unlike other approaches in the literature that employ a tunable reference voltage that can be tuned or controlled after analysing a sample of the generated bitstreams \cite{MemristorPUFFScienceAdv}. Such approaches are usually done in the digital domain analysing a sample of the bitstream after it has been generated and then adapting the reference voltage based on the analysis for future bitstream generation. In our approach, the short-term average of the raw stochastic signal is obtained in the analog domain and in real-time, allowing real-time compensation for any stochastic signal deviation or drift, even prior to bitstream generation. The correction and adaptation of the reference voltage take place before the comparator and without needing any information about the output of the comparator, unlike the other digital approaches that are based on feedback that comes from parts of the system that are after the comparator and require information about the output of the comparator.   

The LPF that generates this adaptive $V_{REF}$ can be implemented using a resistor (R) and a capacitor (C) in a simple RC LPF configuration, as shown in Fig. \ref{Fig 2} (d). The time constant $\tau_{LPF}$ of the RC circuit determines how quickly the reference voltage adapts to changes in $V_{stochastic}$, and is expressed as follows:

\begin{equation}
\tau_{LPF} = RC
\label{eq:tau_LPF}
\end{equation}
where $R$ and $C$ are the resistance and capacitance of the resistor and capacitor in the RC circuit respectively. The cutoff frequency $f_{LPF}$ of the LPF is given by:

\begin{equation}
f_{LPF} =   \frac{1}{2 \pi R C}
\label{eq:f_c}
\end{equation}
   
The selection of $R$ and $C$ depends on how fast the system needs to adapt to fluctuations in $V_{stochastic}$. In a system employing an sMTJ entropy source, $f_{LPF}$ needs to be much lower than the average fluctuation frequency $f_{c}$ of the sMTJ, which can be characterized using the autocorrelation time of magnetization in the sMTJ ($\tau_c$) \cite{ hassan2021quantitative}. This requires the following inequality to be satisfied:

\begin{equation}
\tau_{LPF} > \tau_c
\label{eq:tau_condition}
\end{equation}

Careful design of the sMTJ can allow $\tau_c$ to be across a wide range of timescales from nano-seconds to milli-seconds \cite{borders2019,SubnanosecondFluctuations,NanosecondRTN}, accommodating the needs of different applications, as $\tau_c$ has an impact on the final throughput of the RNG.

\subsection{Post-Processing Circuit: The Trivium Cipher}

After the adaptive digitizer, post-processing is needed to ensure statistical randomness properties of the bit-stream are encryption-grade. For conventional TRNG designs, a commonly employed post-processing circuit is the Trivium Cipher \cite{de2008}. The Trivium cipher is a synchronous stream cipher with an internal state of 288-bits represented by ($s_{1}$, . . . , $s_{288}$) and is intended to produce up to $2^{64}$ key stream bits from an 80-bit secret key and an 80-bit initial value (IV)\cite{montoya2018}. A circuit-level implementation of the Trivium cipher can be seen in the inset of Fig. \ref{Fig 2} (b). The internal state of the cipher is decomposed into three feedback shift registers (FSRs) and coupled with non-linear feedback combinational logic employing only AND and XOR gates.

In Trivium, two out of three FSRs are loaded with the key and IV, while the remaining bits of the internal state are filled in with a constant value. The cipher state is then executed for 4 × 288 = 1152 clock cycles to produce the final bit stream.

\begin{figure*} 
  \centering
  \includegraphics[width=7 in]{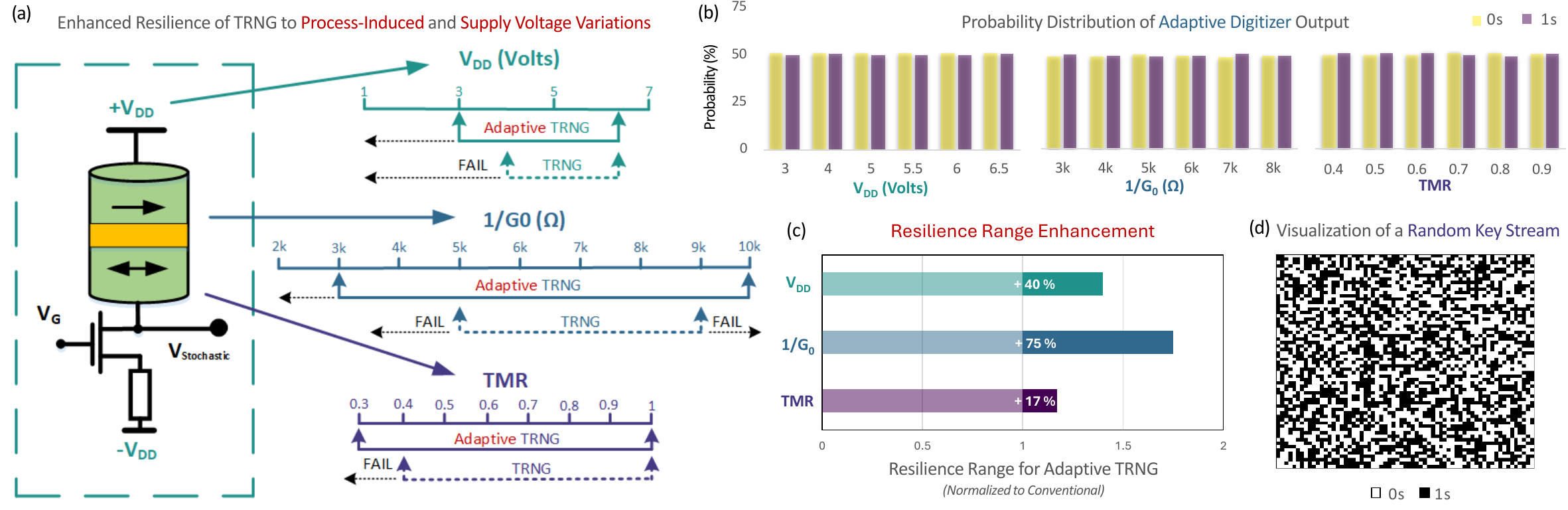}
  \caption{Process-induced and supply-voltage variation analysis for the proposed adaptive TRNG. (a) Ranges of operation with resilience to variation in $V_{DD}$, $G_0$ and $TMR$. (b) Probability of 1’s and 0’s for: varying $V_{DD}$, varying $G_{0}$, and varying $TMR$, from left to right respectively. (c) Resilience range enhancement across the three parameters, $V_{DD}$, $G_0$ and $TMR$, normalized to the resilience range of the conventional design. (d) 2D pattern of a sample generated random bit stream using the presented adaptive TRNG. }
 \label{Fig 4}
\end{figure*}

\subsection{Simulation Results}

Simulation results for the conventional and adaptive TRNG designs are presented in Figs. \ref{Fig 3} (a)-(c) and Figs. \ref{Fig 3} (d)-(f), respectively. Fig. \ref{Fig 3} (a) shows the behavior of the stochastic voltage ($V_{stochastic}$) in the conventional TRNG, while Figs. \ref{Fig 3} (b) and (c) show the response of the digitizer voltage ($V_{Digitizer}$) and the random key stream generated after post-processing (with the conventional Trivium cipher implemented using Cryptool 2.1 (Stable Build) \cite{de2008}), respectively. Fig. \ref{Fig 3} (d) shows the behavior of $V_{stochastic}$ and the adaptive $V_{REF}$ at the output of the low-pass filter in the adaptive TRNG, while Figs. \ref{Fig 3} (e) and (f) show the response of $V_{Digitizer}$ and the generated random key stream, respectively. When compared to the conventional TRNG, the proposed adaptive TRNG dynamically tracks $V_{stochastic}$ and tunes itself to adapt to entropy source variations.

Fig. \ref{Fig 4} (d) depicts a 2D image of a random key stream generated using the proposed adaptive digitizer after post-processing using a Trivium. The random distribution of white and black pixels indicates an unbiased generation of a random ``0/1'' bit-stream. The 2D pattern qualitatively demonstrates the uniform distribution of random bits and validates the design's ability to produce an unbiased bit stream with high-quality randomness.

To quantitatively evaluate the randomness qualities of the generated bit stream, statistical randomness tests are conducted. One commonly employed method is using the NIST test suite (NIST SP 800-22), which is a statistical test suite for random and pseudo-random number generators for cryptographic applications \cite{rukhin2001}, and which will be used in the remainder of the presented study. The suite includes 16 tests, the P-value for each test is used as a measure of randomness and is assessed to determine the test's success.

\subsection{Adaptive TRNG Variation-Resilience Analysis}

In order to investigate the resilience of the adaptive TRNG to variations, simulations of both the conventional and the adaptive TRNG are repeated for a range of process-induced and supply-voltage variations. In addition to supply-voltage variations, the process-induced variations investigated are those that are reflected through variation in two key device parameters: (1) $G_{0}$ and (2) $TMR$. For each case, the simulation was started from a nominal value of the parameter under investigation, then the resulting generated bit stream was tested for randomness using the NIST test suite. For each simulation run a total of 2,398,733 random bits were generated within a period of 0.99 seconds. The adaptive TRNG would be regarded as passing only if the generated bit stream passed all 16 tests. In that case, the obtained p-values were recorded, and then the parameter under test would be incremented with positive increments until it fails by failing at least one test. At that point, the last value where the adaptive TRNG passed all tests would be regarded as the upper limit of the resilience range in that direction. The same procedure would then be repeated in the opposite direction, but with decrements instead of increments, to find the lower limit of the resilience range.

The results of this variation analysis for both the conventional and the proposed adaptive TRNGs are depicted in Fig. \ref{Fig 4} (a) through the visually illustrated variation-resilient ranges, where all tests are passed. The detailed results are also shown in Table \ref{table-1}. For $V_{DD}$ variation analysis, the starting nominal value was 5 V and then the voltage was incremented in 0.5 V increments until the adaptive TRNG failed at 7.0 V, making 6.5 V as the upper limit of the $V_{DD}$ resilience range. The procedure was repeated again with 0.5 V decrements, leading to a 3.0 V lower limit of the $V_{DD}$ resilience range. The same procedure was also conducted for both $G_0$ and $TMR$, and the results are summarized in the visual illustration of Fig. \ref{Fig 4} (a). Moreover, the detailed results are also summarized in Table \ref{table-1}, with the average P-value for each test, across all scenarios where all the tests were passed, is recorded.

\begin{table*}[t]
 \centering
\caption{Statistical Randomness Test Results Under Process-Induced and Supply Voltage Variations (NIST SP 800-22)}
\label{table-1}
\resizebox{\textwidth}{!}{%
\begin{tabular}{||l||cccc|cccc|cccc|cccc||}
\hline
\hline
\rowcolor[HTML]{dadada} \textbf{\scriptsize{Variation Analysis Parameter}}  & & \multicolumn {3} {c|} {\textbf{\scriptsize{$V_{DD}$ Variation}}} & & \multicolumn {3} {c|} {\textbf{\scriptsize{$G_0$ Variation}}} & & \multicolumn {3} {c|} {\textbf{\scriptsize{$TMR$ Variation}}} & & \multicolumn {3} {c||} {\textbf{\scriptsize{$\tau _{c}$ Variation}}} 
\\ \rowcolor[HTML]{dadada}
\textbf{\scriptsize{Test}}  & & \scriptsize{Result} & \scriptsize{P-Value} & \scriptsize{Pass Rate} & & \scriptsize{Result} & \scriptsize{P-Value} & \scriptsize{Pass Rate} & & \scriptsize{Result} & \scriptsize{P-Value} & \scriptsize{Pass Rate}  & & \scriptsize{Result} & \scriptsize{P-Value} & \scriptsize{Pass Rate} 
\\
\hline
\scriptsize{Frequency (Monobits)} & & {\color[HTML]{21918c}\scriptsize{Pass}} & \scriptsize{0.1721} & \scriptsize{8/8} & & {\color[HTML]{21918c}\scriptsize{Pass}} & \scriptsize{0.6424} & \scriptsize{8/8} & & {\color[HTML]{21918c}\scriptsize{Pass}} & \scriptsize{0.8238} & \scriptsize{8/8} & & {\color[HTML]{21918c}\scriptsize{Pass}} & \scriptsize{0.4072} & \scriptsize{2/2} \\

\scriptsize{Frequency within a Block} &&  {\color[HTML]{21918c}\scriptsize{Pass}} & \scriptsize{0.4658} & \scriptsize{8/8} & & {\color[HTML]{21918c}\scriptsize{Pass}} & \scriptsize{0.1161} & \scriptsize{8/8} & & {\color[HTML]{21918c}\scriptsize{Pass}} & \scriptsize{0.2892} & \scriptsize{8/8} & & {\color[HTML]{21918c}\scriptsize{Pass}} & \scriptsize{0.2909} & \scriptsize{2/2} \\

\scriptsize{Runs} & & {\color[HTML]{21918c}\scriptsize{Pass}} & \scriptsize{0.4369} & \scriptsize{8/8} &&  {\color[HTML]{21918c}\scriptsize{Pass}} & \scriptsize{0.7284} & \scriptsize{8/8} &&  {\color[HTML]{21918c}\scriptsize{Pass}} & \scriptsize{0.6554} & \scriptsize{8/8} & & {\color[HTML]{21918c}\scriptsize{Pass}} & \scriptsize{0.5826} & \scriptsize{2/2} \\

\scriptsize{Longest Run of Ones in a Block} & & {\color[HTML]{21918c}\scriptsize{Pass}} & \scriptsize{0.5977} & \scriptsize{8/8} & & {\color[HTML]{21918c}\scriptsize{Pass}} & \scriptsize{0.3048} & \scriptsize{8/8} & & {\color[HTML]{21918c}\scriptsize{Pass}} & \scriptsize{0.3646} & \scriptsize{8/8} & & {\color[HTML]{21918c}\scriptsize{Pass}} & \scriptsize{0.4512} & \scriptsize{2/2} \\

\scriptsize{Binary Matrix Rank} & &  {\color[HTML]{21918c}\scriptsize{Pass}} & \scriptsize{0.4143} & \scriptsize{8/8} & & {\color[HTML]{21918c}\scriptsize{Pass}} & \scriptsize{0.3638} & \scriptsize{8/8} & & {\color[HTML]{21918c}\scriptsize{Pass}} & \scriptsize{0.3338} & \scriptsize{8/8} & & {\color[HTML]{21918c}\scriptsize{Pass}} & \scriptsize{0.3891} & \scriptsize{2/2} \\ 

\scriptsize{Discrete Fourier Transform (Spectral)} & &  {\color[HTML]{21918c}\scriptsize{Pass}} & \scriptsize{0.2287} & \scriptsize{8/8} & & {\color[HTML]{21918c}\scriptsize{Pass}} & \scriptsize{0.4369} & \scriptsize{8/8} & & {\color[HTML]{21918c}\scriptsize{Pass}} & \scriptsize{0.7255} & \scriptsize{8/8} & & {\color[HTML]{21918c}\scriptsize{Pass}} & \scriptsize{0.3328} & \scriptsize{2/2} \\ 

\scriptsize{Non-Overlapping Template Matching} & & {\color[HTML]{21918c}\scriptsize{Pass}} & \scriptsize{0.6114} & \scriptsize{8/8} & & {\color[HTML]{21918c}\scriptsize{Pass}} & \scriptsize{0.1671} & \scriptsize{8/8} & & {\color[HTML]{21918c}\scriptsize{Pass}} & \scriptsize{0.2771} & \scriptsize{8/8} & & {\color[HTML]{21918c}\scriptsize{Pass}} & \scriptsize{0.3892} & \scriptsize{2/2}\\ 

\scriptsize{Overlapping Template Matching} & & {\color[HTML]{21918c}\scriptsize{Pass}} & \scriptsize{0.7457} & \scriptsize{8/8} & & {\color[HTML]{21918c}\scriptsize{Pass}} & \scriptsize{0.3901} & \scriptsize{8/8} & & {\color[HTML]{21918c}\scriptsize{Pass}} & \scriptsize{0.3601} & \scriptsize{8/8}& & {\color[HTML]{21918c}\scriptsize{Pass}} & \scriptsize{0.5679} & \scriptsize{2/2} \\ 

\scriptsize{Maurer's "Universal Statistical"} & & {\color[HTML]{21918c}\scriptsize{Pass}} & \scriptsize{0.6514} & \scriptsize{8/8} & & {\color[HTML]{21918c}\scriptsize{Pass}} & \scriptsize{0.5845} & \scriptsize{8/8} & & {\color[HTML]{21918c}\scriptsize{Pass}} & \scriptsize{0.4523} & \scriptsize{8/8} & & {\color[HTML]{21918c}\scriptsize{Pass}} & \scriptsize{0.6179} & \scriptsize{2/2}\\ 

\scriptsize{Linear Complexity} & & {\color[HTML]{21918c}\scriptsize{Pass}} & \scriptsize{0.5402} & \scriptsize{8/8} & & {\color[HTML]{21918c}\scriptsize{Pass}} & \scriptsize{0.7001} & \scriptsize{8/8} & & {\color[HTML]{21918c}\scriptsize{Pass}} & \scriptsize{0.1062} & \scriptsize{8/8} & & {\color[HTML]{21918c}\scriptsize{Pass}} & \scriptsize{0.6201} & \scriptsize{2/2} \\ 

\scriptsize{Serial} & & {\color[HTML]{21918c}\scriptsize{Pass}} & \scriptsize{0.7153} & \scriptsize{8/8} & & {\color[HTML]{21918c}\scriptsize{Pass}} & \scriptsize{0.0745} & \scriptsize{8/8} & & {\color[HTML]{21918c}\scriptsize{Pass}} & \scriptsize{0.7624} & \scriptsize{8/8} & & {\color[HTML]{21918c}\scriptsize{Pass}} & \scriptsize{0.5183} & \scriptsize{2/2} \\ 

\scriptsize{Approximate Entropy} & & {\color[HTML]{21918c}\scriptsize{Pass}} & \scriptsize{0.2519} & \scriptsize{8/8} & &  {\color[HTML]{21918c}\scriptsize{Pass}} & \scriptsize{0.3214} & \scriptsize{8/8} & & {\color[HTML]{21918c}\scriptsize{Pass}} & \scriptsize{0.5214} & \scriptsize{8/8} & & {\color[HTML]{21918c}\scriptsize{Pass}} & \scriptsize{0.4751} & \scriptsize{2/2} \\ 

\scriptsize{Cumulative Sums (Forward)} & & {\color[HTML]{21918c}\scriptsize{Pass}} & \scriptsize{0.3192} & \scriptsize{8/8} & & {\color[HTML]{21918c}\scriptsize{Pass}} & \scriptsize{0.6984} & \scriptsize{8/8} & & {\color[HTML]{21918c}\scriptsize{Pass}} & \scriptsize{0.6109} & \scriptsize{8/8} & & {\color[HTML]{21918c}\scriptsize{Pass}} & \scriptsize{0.3419} & \scriptsize{2/2} \\ 

\scriptsize{Cumulative Sums (Reverse)} & & {\color[HTML]{21918c}\scriptsize{Pass}} & \scriptsize{0.1762} & \scriptsize{8/8} & & {\color[HTML]{21918c}\scriptsize{Pass}} & \scriptsize{0.3647} & \scriptsize{8/8} & & {\color[HTML]{21918c}\scriptsize{Pass}} & \scriptsize{0.5274} & \scriptsize{8/8} & & {\color[HTML]{21918c}\scriptsize{Pass}} & \scriptsize{0.4626} & \scriptsize{2/2} \\ 

\scriptsize{Random Excursions} & & {\color[HTML]{21918c}\scriptsize{Pass}} & \scriptsize{0.8595} & \scriptsize{8/8} & &  {\color[HTML]{21918c}\scriptsize{Pass}} & \scriptsize{0.7491} & \scriptsize{8/8} & &  {\color[HTML]{21918c}\scriptsize{Pass}} & \scriptsize{0.5942} & \scriptsize{8/8} & & {\color[HTML]{21918c}\scriptsize{Pass}} & \scriptsize{0.5456} & \scriptsize{2/2} \\ 

\scriptsize{Random Excursions Variant} & & {\color[HTML]{21918c}\scriptsize{Pass}} & \scriptsize{0.5971} & \scriptsize{8/8} & & {\color[HTML]{21918c}\scriptsize{Pass}} & \scriptsize{0.4941} & \scriptsize{8/8} & & {\color[HTML]{21918c}\scriptsize{Pass}} & \scriptsize{0.6424} & \scriptsize{8/8} & & {\color[HTML]{21918c}\scriptsize{Pass}} & \scriptsize{0.5947} & \scriptsize{2/2} \\  
\hline
\hline

\end{tabular}}
\end{table*}

The results of Fig. \ref{Fig 4} and Table \ref{table-1} demonstrate the enhanced resilience of the adaptive TRNG to variations in $V_{DD}$, $G_{0}$ and $TMR$ compared to a conventional TRNG, reaching up to an enhancement in the variation-resilience range of 40 \%, 75 \% and 17 \%, respectively, as illustrated in Fig. \ref{Fig 4} (c). Fig. \ref{Fig 4} (b), illustrates the percentage of 1s and 0s for six different cases of varying $V_{DD}$, $G_{0}$, and $TMR$, respectively. As is evident in all test cases, the percentage of 1s and 0s is close to an ideal 50 \% each, thus confirming that the bit patterns are unbiased.

Furthermore, to test the system's ability of operation at different throughput ranges, we conducted the tests for two cases of autocorrelation time ($\tau_c$) variation, one with $\tau_c$ = 25 $\mu$s and the other with $\tau_c$ = 2.5 $\mu$s, giving rise to an overall system throughput of 5 Mbps and 182.5 Mbps, respectively. These results are also presented in Table \ref{table-1}. It would be important to highlight here that the presented adaptive RNG architecture is not limited to these specific values of throughput, indeed it may be able to support higher throughput, however, the throughput will depend on the speed of fluctuation in the entropy source (sMTJ in this case). The used values of the autocorrelation time ($\tau_c$) for the stochastic MTJs in our simulations are moderate values that lie in the range of abundantly reported experimental values of $\tau_c$ on the order of microseconds. Nonetheless, a few experimental reports have indeed reported much smaller values on the order of nanoseconds \cite{SubnanosecondFluctuations,NanosecondRTN}, which may result in much higher throughput, however, these experimental reports are for very early stage nanoscale technologies that are yet to reach an acceptable level of technology maturity for mass production. 

\begin{figure} [t]
   \centering
  \includegraphics[width=\columnwidth]{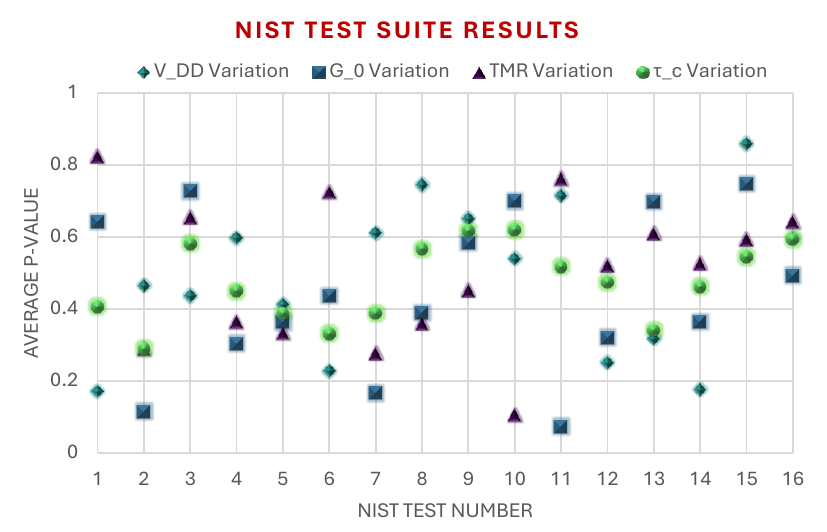}
  \caption{A scatter plot as a visual illustration of the NIST Test Suite results for the variation analysis. Average P-values across the resilience range to variation in $V_{DD}$, $G_0$, $TMR$ and $\tau_{c}$ for the adaptive TRNG.}
  \label{Fig 5}
\end{figure}

Fig. \ref{Fig 5} shows a visual illustrative scatter plot of the obtained average P-values from the NIST test suite in the presence of variations, demonstrating that they all fall well within the acceptable range, further highlighting the resilient nature of the adaptive TRNG.

\section{Adaptive RNG with a Pseudo-Random Entropy Source}

While TRNGs with physically driven naturally random entropy sources provide very high quality true-randomness desired for encryption, the advanced device technologies needed for these TRNGs are not accessible to all IoT device designers, nor are they always applicable in IoT applications that do not justify the high cost of custom designing chips with these technologies. In this section, we show how our adaptive RNG can address this issue by providing a low-cost highly accessible route to embedded encryption in IoT devices. To achieve this, we employ a low-quality pseudo-random entropy source, an LFSR, which can be viewed as an emulation of the sMTJ but with pseudo randomness. An LFSR is readily implementable on entry-level FPGA prototyping kits, and its design is a well-known text-book design. However, an LFSR on its own does not meet the randomness requirements of encryption applications. Using our adaptive RNG system, we show how the use of the adaptive digitizer with simplified post-processing can increase the entropy of the LFSR, making it encryption-grade.

\subsection{Enhancing LFSR Randomness with The Adaptive Digitizer}

In order to enhance the entropy of the pseudo-random source, the digital output from the LFSR is converted to an analog signal serving as $V_{Stochastic}$, and is then fed into the adaptive digitizer that converts it back to digital. However, here the comparison is against the adaptive moving average - $V_{REF}$ - of this stochastic signal, and hence the generated output bit stream is different than the original bit stream obtained from the LFSR. This output of the adaptive digitizer will not be the same as the inputted digital bit stream from the LFSR, but will have higher entropy due to the conversion and subsequent adaptive digitization processes that infuses more noise (randomness) into the signal. This output will then undergo post-processing to produce a digital random bit stream with randomness qualities that fulfil the basic requirements of embedded cryptography applications.

\begin{figure} [t]
   \centering
  \includegraphics[width=\columnwidth]{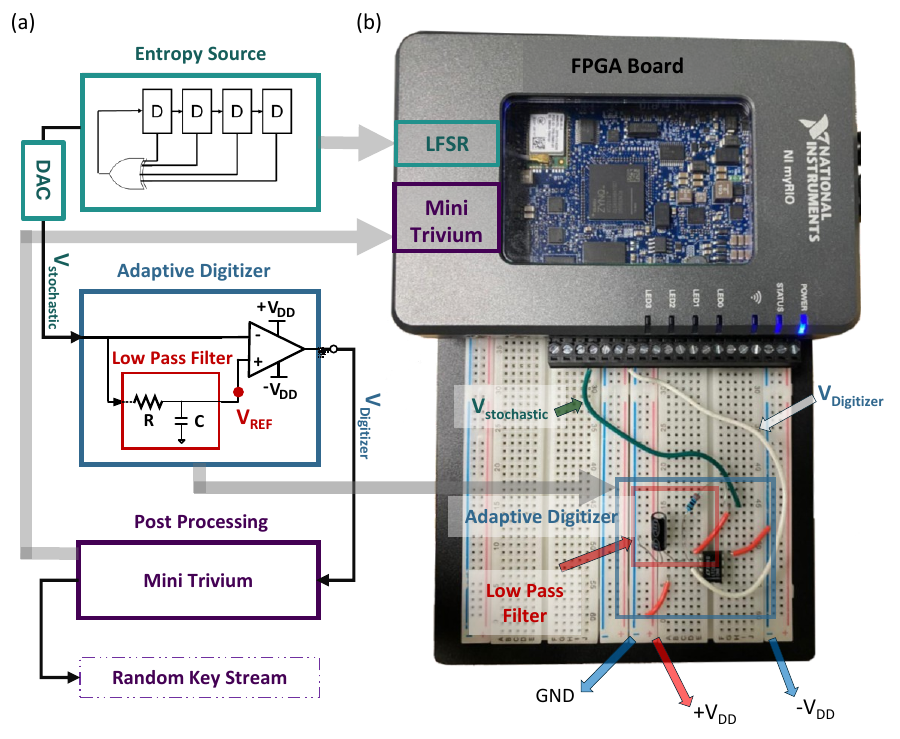}
  \caption{(a) Schematic diagram of system blocks and interconnections. (b) Experimental implementation of the adaptive RNG system using discrete electronic components and an FPGA board.}
  \label{Fig 6}
\end{figure}

\subsection{Compact Post-Processing: The Mini Trivium} 

Although the Trivium Cipher is regarded as a relatively low area hardware cipher, it still requires thousands of logic gates for implementation and can impose relatively heavy power consumption levels for energy-constrained IoT devices, especially when conducting continuous real-time data encryption; essential in remote sensing. In order to further reduce the required hardware resources for embedded encryption in resource-constrained IoT devices, we also present a modified lower-hardware-cost version of the Trivium cipher used for post-processing, and name it here as a Mini Trivium. 

Due to the enhanced randomness qualities of the generated bit stream from the adaptive digitizer, it is possible to achieve adequate post-processing using a circuit with reduced complexity. Accordingly, we employ the simplified Mini Trivium cipher, whose circuit implementation is shown in Fig. \ref{Fig 2} (d).

The Mini Trivium circuit simplifies the original Trivium Cipher design by reducing its three Nonlinear Feedback Shift Registers (NFSRs) into a single 19-bit shift register, thus reducing complexity, without compromising core cryptographic properties. By strategically selecting taps ${t_1}$ to ${t_{10}}$ and reducing core logic gates down to 10 gates only, it achieves faster state updates and key stream generation through streamlined XOR and AND operations. The shorter register length and reduced feedback paths enable quicker initialization cycles compared to Trivium’s 288-bit state. Despite fewer bits, non-linearity is preserved via carefully placed gates, balancing security and efficiency.

\begin{figure} [t]
   \centering
  \includegraphics[width=\columnwidth]{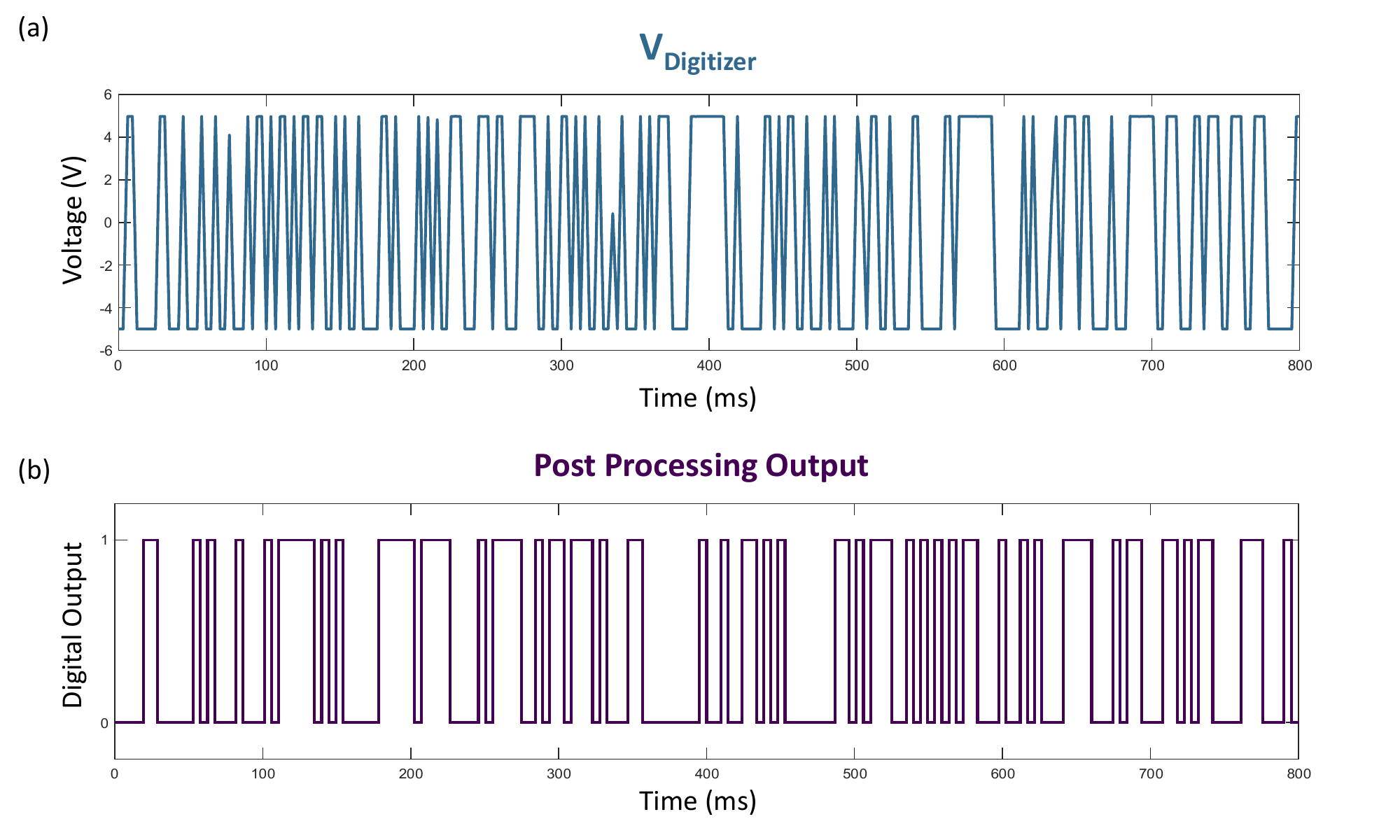}
  \caption{Measured results from the FPGA experiments, showing (a) the output of the adaptive digitizer and (b) the corresponding generated random bit stream after post-processing with the Mini Trivium.}
  \label{Fig 7}
\end{figure}

\begin{table*}[t]
 \centering
\caption{Comparison Between Different RNG Types (NIST SP 800-22 Test Suite)}
\label{table-2}
\begin{tabular}{||l||cccc|cccc|cccc||}
\hline
\hline
\rowcolor[HTML]{dadada} \textbf{\scriptsize{System Under Test}}  & & \multicolumn {3} {c|} {\textbf{\scriptsize{LFSR ONLY*}}} & & \multicolumn {3} {c|} {\textbf{\scriptsize{Adaptive RNG (LFSR)**}}} & & \multicolumn {3} {c||} {\textbf{\scriptsize{Adaptive TRNG (sMTJ)***}}} 
\\ \rowcolor[HTML]{dadada}
\textbf{\scriptsize{Test}}  & & \scriptsize{Result} & \scriptsize{P-Value} & \scriptsize{Pass Rate} & & \scriptsize{Result} & \scriptsize{P-Value} & \scriptsize{Pass Rate} & & \scriptsize{Result} & \scriptsize{P-Value} & \scriptsize{Pass Rate} 
\\
\hline
\scriptsize{Frequency (Monobits)}  & & {\color[HTML]{440154}\scriptsize{Fail}} & \scriptsize{0.000} & \scriptsize{0/10} & &{\color[HTML]{21918c}\scriptsize{Pass}} & \scriptsize{0.414} & \scriptsize{10/10} & &{\color[HTML]{21918c}\scriptsize{Pass}} & \scriptsize{0.799} & \scriptsize{10/10} \\ 
\scriptsize{Frequency within a Block}  & & {\color[HTML]{440154}\scriptsize{Fail}} & \scriptsize{0.000} & \scriptsize{0/10} & &{\color[HTML]{21918c}\scriptsize{Pass}} & \scriptsize{0.479} & \scriptsize{10/10} & &{\color[HTML]{21918c}\scriptsize{Pass}} & \scriptsize{0.851} & \scriptsize{10/10} \\ 
\scriptsize{Runs}  & & {\color[HTML]{440154}\scriptsize{Fail}} & \scriptsize{0.000} & \scriptsize{0/10} & &{\color[HTML]{21918c}\scriptsize{Pass}} & \scriptsize{0.728} & \scriptsize{10/10} & &{\color[HTML]{21918c}\scriptsize{Pass}} & \scriptsize{0.961} & \scriptsize{10/10} \\ 
\scriptsize{Longest Run of Ones in a Block}  & & {\color[HTML]{440154}\scriptsize{Fail}} & \scriptsize{0.000} & \scriptsize{0/10} & &{\color[HTML]{21918c}\scriptsize{Pass}} & \scriptsize{0.946} & \scriptsize{10/10} & &{\color[HTML]{21918c}\scriptsize{Pass}} & \scriptsize{0.727} & \scriptsize{10/10} \\ 
\scriptsize{Binary Matrix Rank} &  &{\color[HTML]{440154}\scriptsize{Fail}} & \scriptsize{0.000} & \scriptsize{0/10} & &{\color[HTML]{21918c}\scriptsize{Pass}} & \scriptsize{0.881} & \scriptsize{10/10} & &{\color[HTML]{21918c}\scriptsize{Pass}} & \scriptsize{0.333} & \scriptsize{10/10} \\ 
\scriptsize{Discrete Fourier Transform (Spectral)} &  & {\color[HTML]{440154}\scriptsize{Fail}} & \scriptsize{0.000} & \scriptsize{0/10} & &{\color[HTML]{21918c}\scriptsize{Pass}} & \scriptsize{0.223} & \scriptsize{10/10} & &{\color[HTML]{21918c}\scriptsize{Pass}} & \scriptsize{0.569} & \scriptsize{10/10} \\ 
\scriptsize{Non-Overlapping Template Matching} &  & {\color[HTML]{440154}\scriptsize{Fail}} & \scriptsize{0.000} & \scriptsize{0/10} & &{\color[HTML]{21918c}\scriptsize{Pass}} & \scriptsize{0.274} & \scriptsize{10/10} & &{\color[HTML]{21918c}\scriptsize{Pass}} & \scriptsize{0.619} & \scriptsize{10/10} \\ 
\scriptsize{Overlapping Template Matching} &  & {\color[HTML]{440154}\scriptsize{Fail}} & \scriptsize{0.000} & \scriptsize{0/10} & &{\color[HTML]{21918c}\scriptsize{Pass}} & \scriptsize{0.219} & \scriptsize{10/10} & &{\color[HTML]{21918c}\scriptsize{Pass}} & \scriptsize{0.662} & \scriptsize{10/10} \\ 
\scriptsize{Maurer's "Universal Statistical"} &  & {\color[HTML]{440154}\scriptsize{Fail}} & \scriptsize{0.000} & \scriptsize{0/10} & &{\color[HTML]{21918c}\scriptsize{Pass}} & \scriptsize{0.434} & \scriptsize{10/10} & &{\color[HTML]{21918c}\scriptsize{Pass}} & \scriptsize{0.551} & \scriptsize{10/10} \\ 
\scriptsize{Linear Complexity} &  &{\color[HTML]{440154}\scriptsize{Fail}} & \scriptsize{0.000} & \scriptsize{0/10} & &{\color[HTML]{21918c}\scriptsize{Pass}} & \scriptsize{0.754} & \scriptsize{10/10} & &{\color[HTML]{21918c}\scriptsize{Pass}} & \scriptsize{0.314} & \scriptsize{10/10} \\ 
\scriptsize{Serial}  & & {\color[HTML]{440154}\scriptsize{Fail}} & \scriptsize{0.000} & \scriptsize{0/10} & &{\color[HTML]{21918c}\scriptsize{Pass}} & \scriptsize{0.831} & \scriptsize{10/10} & &{\color[HTML]{21918c}\scriptsize{Pass}} & \scriptsize{0.381} & \scriptsize{10/10} \\ 
\scriptsize{Approximate Entropy} &  &{\color[HTML]{440154}\scriptsize{Fail}} & \scriptsize{0.000} & \scriptsize{0/10} & &{\color[HTML]{21918c}\scriptsize{Pass}} & \scriptsize{0.556} & \scriptsize{10/10} & &{\color[HTML]{21918c}\scriptsize{Pass}} & \scriptsize{0.445} & \scriptsize{10/10} \\ 
\scriptsize{Cumulative Sums (Forward)} &  &{\color[HTML]{21918c}\scriptsize{Pass}} & \scriptsize{1.000} & \scriptsize{10/10} & &{\color[HTML]{21918c}\scriptsize{Pass}} & \scriptsize{0.782} & \scriptsize{10/10} & &{\color[HTML]{21918c}\scriptsize{Pass}} & \scriptsize{0.652} & \scriptsize{10/10} \\ 
\scriptsize{Cumulative Sums (Reverse)} &  &{\color[HTML]{21918c}\scriptsize{Pass}} & \scriptsize{1.000} & \scriptsize{10/10} & &{\color[HTML]{21918c}\scriptsize{Pass}} & \scriptsize{0.309} & \scriptsize{10/10} & &{\color[HTML]{21918c}\scriptsize{Pass}} & \scriptsize{0.625} & \scriptsize{10/10} \\ 
\scriptsize{Random Excursions} &  &{\color[HTML]{21918c}\scriptsize{Pass}} & \scriptsize{0.849} & \scriptsize{10/10} & &{\color[HTML]{21918c}\scriptsize{Pass}} & \scriptsize{0.521} & \scriptsize{10/10} & &{\color[HTML]{21918c}\scriptsize{Pass}} & \scriptsize{0.885} & \scriptsize{10/10} \\ 
\scriptsize{Random Excursions Variant} &  &{\color[HTML]{21918c}\scriptsize{Pass}} & \scriptsize{0.617} & \scriptsize{10/10} & &{\color[HTML]{21918c}\scriptsize{Pass}} & \scriptsize{0.778} & \scriptsize{10/10} & &{\color[HTML]{21918c}\scriptsize{Pass}} & \scriptsize{0.629} & \scriptsize{10/10} \\  
\hline
\hline

\multicolumn {3} {l} {\scriptsize{*Experiment: entropy source: LFSR.}}        \\                 
\multicolumn {5} {l} {\scriptsize{**Experiment: entropy source: LFSR + DAC, post-processing: Mini Trivium.}}                       \\  
\multicolumn {7} {l} {\scriptsize{***Simulation: entropy source: sMTJ and associated circuitry, post-processing: Trivium.}}                       \\  

\end{tabular}
\end{table*}

\subsection{Experimental Implementation and Results} 

To test the overall proposed adaptive RNG system with the LFSR and the Mini-Trivium, the design was experimentally implemented using discrete electronic components and an FPGA board (NI myRIO-1900), as depicted in Fig. \ref{Fig 6} (b). The schematic of Fig. \ref{Fig 6} (a) depicts the complete system implementation, showing the connections between various components of the system. The combination of the linear feedback shift register (LFSR) along with the digital-to-analog converter (DAC) is employed as the entropy source that generates an analog stochastic signal ($V_{Stochastic}$), similar to the signal produced by the sMTJ and its associated circuitry presented in the previous section. Once generated, $V_{Stochastic}$ is then fed into both the LPF and and the comparator analog circuits within the adaptive digitizer. The LPF and the comparator of the adaptive digitizer were implemented using off-the-shelf discrete components on a breadboard, nonetheless, they can also be implemented using integrated components.

A sample output from the adaptive digitizer is shown in Fig. \ref{Fig 7} (a). As shown in the figure, the output looks like a non-ideal digital signal with finite slopes, and this is due to the limited slew rate (SR) of the used operational amplifier (OpAmp) for the comparator \cite{Mohammed_2025}. To obtain more ideal digital waveforms, a comparator with a higher SR can be employed or designed. This output from the adaptive digitizer - obtained using the breadboard-implemented analog circuit shown in Fig. \ref{Fig 6} (b) - is then passed through the Mini Trivium cipher for post-processing. Fig. \ref{Fig 7} (b) shows the random bit stream generated from the output of the adaptive digitizer that was shown in Fig. \ref{Fig 7} (a) after post-processing. The compact Mini Trivium post-processing circuit was implemented using the FPGA unit. The FPGA unit was programmed and tested using a LABVIEW environment.

\begin{figure} [t]
   \centering
  \includegraphics[width=\columnwidth]{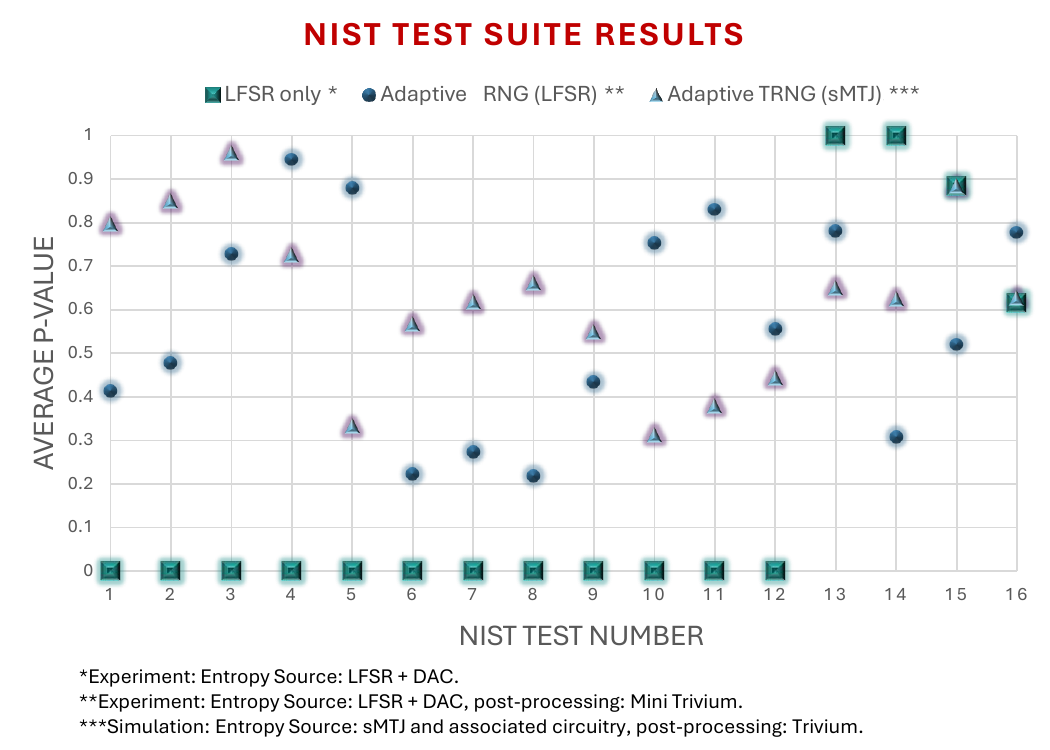}
  \caption{NIST Test Suite results comparing a conventional LFSR with both an LFSR-based adaptive RNG and an sMTJ-based adaptive TRNG}
  \label{Fig 8}
\end{figure}

In order to investigate the enhancement that the adaptive RNG system provides to the raw LFSR entropy source, when running any experiment both: (1) the raw bit stream generated by the LFSR only, and (2) the final generated random bit stream after the adaptive digitizer and post-processing were recorded. The system was operated to generate a random bit stream of 1,670,000 bits in each experiment. This run was repeated 10 independent times providing 10 independent experiments. Each time the experiment was conducted, the generated raw random bit stream from the LFSR only and the final random bit stream generated by the adaptive RNG were each tested individually using the NIST test suite for randomness. A similar set of simulation runs were also conducted on the sMTJ-based adaptive TRNG system for comparison purposes. The results for all three systems: (1) LFSR only, (2) LFSR-based adapative RNG, and (3) sMTJ-based adaptive TRNG are summarized in Table \ref{table-2}. Moreover, the results for the average P-values for each individual NIST test for all the three systems, are presented visually through the scatter plot of Fig. \ref{Fig 8}.

\begin{table*}[t]
  \centering
  \caption{Performance Comparison of State-of-The-Art TRNGs}
  \label{table-comparison}
  
  {\scriptsize
  \setlength{\tabcolsep}{2.2pt}
  \renewcommand{\arraystretch}{1.6}
  \begin{tabular}{||w{l}{4.0cm}||c|c|c|c|c|c|c|c|c||}
    \hline
    \hline
    \rowcolor[HTML]{dadada}
    \textbf{Reference and Year} &
    \cite{cao2021new} &
    \cite{luo20232} &
    \cite{bae20163} &
    \cite{kim2019nano} &
    \cite{arumi2023true} &
    \cite{fu2023rhs} &
    \cite{zahoor2024novel} &
    \cite{koh2025closed} &
    \textbf{This} \\
    
    \rowcolor[HTML]{dadada}
    \textbf{Parameter} &
    2022 &
    2023 &
    2017 &
    2019 &
    2022 &
    2023 &
    2024 &
    2025 &
    \textbf{Work} \\
    \hline
    Entropy Source Technology &
    CMOS &
    CMOS &
    CMOS &
    RRAM &
    RRAM &
    MTJ &
    MTJ &
    sMTJ &
    sMTJ \\
    \hline
    
    Entropy Source Mechanism &
    \makecell[c]{Oscillator\\Jitter} &
    \makecell[c]{Oscillator\\Jitter} &
    \makecell[c]{Metastability\\and Jitter} & 
    \makecell[c]{Random\\Telegraph\\Noise} &
    \makecell[c]{Intra-\\Device\\Switching} &
    \makecell[c]{Static\\Stochastic\\Switching} &
    \makecell[c]{Stochastic\\Initialization\\Switching} &
    \makecell[c]{Static\\Stochastic\\Switching} &
    \makecell[c]{Dynamic\\Stochastic\\Switching} \\
    \hline
    
    Operating Voltage ($V_{DD}$) &
    $1.2$ V &
    $1.1$ V &
    $1.2$ V &
    --  &
    $0.7$ V &
    --  &
    --  &
    --  &
    $5.0$ V \\
    \hline

    Switching Voltage ($V_{SET}$) &
    -- &
    -- &
    -- &
    $1.0$ V &
    -- &
    --&
    $2.0$ V &
    -- &
    -- \\
    \hline
    
    Throughput (Mbps) &
    52 &
    53 &
    3000 &
    40 &
    -- &
    303 &
    -- &
    -- &
    5--182$^{*}$ \\
    \hline
    
    NIST Tests Pass Rate &
    15/15 &
    15/15 &
    15/15 &
    15/15 &
    15/15 &
    14/14 &
    16/16 &
    15/15 &
    16/16 \\
    \hline
    
    Variation Resilient &
    {\color[HTML]{21918c}\scriptsize{Yes}} &
    {\color[HTML]{21918c}\scriptsize{Yes}} &
    {\color[HTML]{21918c}\scriptsize{Yes}} &
    {\color[HTML]{21918c}\scriptsize{Yes}} &
    {\color[HTML]{440154}\scriptsize{No}} &
    {\color[HTML]{21918c}\scriptsize{Yes}} &
    {\color[HTML]{440154}\scriptsize{No}} &
    {\color[HTML]{21918c}\scriptsize{Yes}} &
    \textbf{{\color[HTML]{21918c}\scriptsize{Yes}}} \\
    \hline
    
    Post-Processing Included&
    {\color[HTML]{440154}\scriptsize{No}} &
    {\color[HTML]{21918c}\scriptsize{Yes}} &
    {\color[HTML]{440154}\scriptsize{No}} &
    {\color[HTML]{21918c}\scriptsize{Yes}} &
    {\color[HTML]{440154}\scriptsize{No}} &
    {\color[HTML]{440154}\scriptsize{No}} &
    {\color[HTML]{21918c}\scriptsize{Yes}} &
    {\color[HTML]{440154}\scriptsize{No}} &
    \textbf{{\color[HTML]{21918c}\scriptsize{Yes}}} \\
    \hline
    
    Off-Shelf Implementation &
    -- &
    -- &
    -- &
    -- &
    -- &
    -- &
    -- &
    -- &
    \textbf{{\color[HTML]{21918c}\scriptsize{Yes}}} \\
    \hline
    
    Entropy Source Generality &
    -- &
    -- &
    -- &
    -- &
    -- &
    -- &
    -- &
    -- &
    \textbf{{\color[HTML]{21918c}\scriptsize{Yes}}} \\
    \hline
    \hline
  \end{tabular}
  }

   \vspace{0.05cm}
  {\scriptsize{CMOS: Complementary Metal Oxide Semiconductor | RRAM: Resistive Random Access Memory | MTJ: Magnetic Tunnel Junction
  \\
  $^{*}$Based on reported results in this paper but in principle can go to much higher throughput (up to Gbps) with faster stochastic MTJs \cite{SubnanosecondFluctuations,NanosecondRTN}.}}
  \end{table*}

As the results of Table \ref{table-2} show, the LFSR on its own fails 12 out 16 tests consistently in all 10 experiments, confirming that the LFSR on its own is a low-quality PRNG. On the other hand, when the same random bit stream generated by the LFSR goes through the adaptive RNG system (namely the combination of the DAC, adaptive digitizer and the Mini Trivium cipher) then the resulting random bit stream shows excellent statistical randomness properties, consistently passing all 16 tests during all conducted experiments. Not only do the random bit streams from the LFSR-based adaptive RNG pass the tests, but they do so with excellent P-values consistently above 0.2. The obtained P-values are indeed comparable to the results obtained for the sMTJ-based adaptive TRNG system at nominal conditions, as shown quantitatively through both Table \ref{table-2} and the visual illustration of Fig. \ref{Fig 8}. The experimental results confirm that the LFSR-based adaptive RNG implementation using off-the-shelf components results in the generation of high quality random numbers with properties suitable for embedded encryption.

Finally, to benchmark our presented adaptive RNG with other implementations, a performance comparison of our adaptive RNG with the existing state-of-the art TRNG designs based on CMOS, RRAM and MTJ technologies is presented in Table \ref{table-comparison}. The presented adaptive RNG offers a wide - relatively high - throughput range of 5 Mbps to 182 Mbps ($\tau_c$ of 25 $\mu$s - 2.5 $\mu$s respectively), and a consistent 100\% NIST test pass rate.  It is important to mention here, that our approach does not need voltage pulsing as required in other designs, and also our design is implementable using  off-shelf component implementations. It is important to also highlight the generality of our adaptive RNG concept beyond MTJs only, as it can be implemented using various entropy sources (e.g. CMOS, RRAM, MTJ), as demonstrated through our LFSR implementation in the FPGA experiments.

\section{Conclusion}

In this work, we demonstrated an adaptive RNG for variation-resilient extraction of random numbers from entropy sources based on emerging device technologies. Two demonstrations were presented, one simulation-based and one experimental, with the first employing an sMTJ-based entropy source, which was comprehensively evaluated using MATLAB + SPICE simulations, and the other employing an LFSR-based pseudo random entropy source that emulates the true-random source, which was experimentally implemented using discrete electronic components and an FPGA. 

A key component of the adaptive RNG system was its adaptive digitizer, which employed a low-pass filter to generate an adaptive dynamically changing short-term-average reference voltage that tracks any deviation or drift in the generated stochastic signal from the entropy source, enabling variation-resilience and allowing extraction of unbiased random bit streams with high entropy. The presented adaptive RNG consistently passed all NIST SP 800-22 tests, and demonstrated a throughput in the range 5-182 Mbps, while maintaining high randomness qualities. Although the NIST test suite is the commonly used framework for evaluating randomness qualities of random bit streams generated using TRNGs, the presented analysis can be further expanded in the future to include other methods as well, such as the Dieharder \cite{marsaglia2002some}, TestU01 \cite{l2007testu01,foreman2024statistical,foreman2023practical} and PractRand \cite{sleem2020testu01}, which can be useful for more comprehensive assessment of the randomness qualities of TRNGs for mission critical and highly sensitive encryption applications. The presented adaptive RNG has great potential for use in embedded encryption within IoT devices, making data encryption accessible locally.

\section*{Acknowledgment}

Authors thank Mahmood A. Mohammed for feedback. Part of the work of F. Zahoor and F. Al-Dirini was done at KFUPM. Authors acknowledge support of the Natural Sciences and Engineering Research Council of Canada (NSERC) [Grant number: RGPIN-2023-03743], and partial support of KFUPM [Grant number: INAM2306].

\bibliographystyle{modified-apsrev4-2}
\bibliography{refs}

@article{dubovskiy2024one,
  title={One Trillion True Random Bits Generated with a Field Programmable Gate Array Actuated Magnetic Tunnel Junction},
  author={Dubovskiy, Andre and Criss, Troy and El Valli, Ahmed Sidi and Rehm, Laura and Kent, Andrew D and Haas, Andrew},
  journal={	IEEE Magnetics Letters
},
volume={15},
  number={1},
  pages={1--4},
  year={2024},
 publisher={IEEE},
doi={10.1109/LMAG.2024.3416091}
}

@article{qu2018variation,
  title={Variation-resilient true random number generators based on multiple STT-MTJs},
  author={Qu, Yuanzhuo and Cockburn, Bruce F and Huang, Zhe and Cai, Hao and Zhang, Yue and Zhao, Weisheng and Han, Jie},
  journal={IEEE 	 Transaction on Nanotechnology
},
  volume={17},
  number={6},
  pages={1270--1281},
  year={2018},
  publisher={IEEE},
doi={10.1109/TNANO.2018.2873970}
}

@article{camsari2017,
  title={Implementing p-bits with embedded MTJ},
  author={Camsari, Kerem Yunus and Salahuddin, Sayeef and Datta, Supriyo},
  journal={IEEE Electron Device Letters},
  volume={38},
  number={12},
  pages={1767--1770},
  year={2017},
  publisher={IEEE},
doi={10.1109/LED.2017.2768321}
}

@article{borders2019,
  doi = {10.1038/s41586-019-1557-9},
  url = {https://doi.org/10.1038/s41586-019-1557-9},
  year = {2019},
  month = sep,
  publisher = {Springer Science and Business Media {LLC}},
  volume = {573},
  number = {7774},
  pages = {390--393},
  author = {William A. Borders and Ahmed Z. Pervaiz and Shunsuke Fukami and Kerem Y. Camsari and Hideo Ohno and Supriyo Datta},
  title = {Integer factorization using stochastic magnetic tunnel junctions},
  journal = {Nature}
}

@inproceedings{bunaiyan2022mtj,
  title={MTJ-Based p-Bit Designs for Enhanced Tunability},
  author={Bunaiyan, Saleh and Al-Dirini, Feras},
  booktitle={2022 IEEE Nanotechnology Materials and Devices Conference (NMDC)},
  pages={14--16},
  year={2022},
  organization={IEEE},
doi={10.1109/NMDC46933.2022.10052369}
}

@incollection{de2008,
  title={Trivium},
  author={De Canniere, Christophe and Preneel, Bart},
  booktitle={New Stream Cipher Designs: The eSTREAM Finalists},
  pages={244--266},
  year={2008},
  publisher={Springer},
doi={10.1007/978-3-540-68351-3}
}

@inproceedings{montoya2018,
  title={Energy-efficient masking of the trivium stream cipher},
  author={Montoya, Maxime and Hiscock, Thomas and Bacles-Min, Simone and Molnos, Anca and Fournier, Jacques JA},
  booktitle={	25th IEEE Int. Conf. on Electronics, Circuits and Syst. (ICECS), 2018, pp. 393--396
},
  pages={393--396},
  year={2018},
doi={10.1109/ICECS.2018.8617892}
}

@article{tian2009,
  title={On the design of Trivium},
  author={Tian, Yun and Chen, Gongliang and Li, Jianhua},
  journal={Cryptology ePrint Archive},
  year={2009},
  url = {https://eprint.iacr.org/2009/431}
}

@inproceedings{bunaiyan2021real,
  title={Real-time analog event-detection for event-based synchronous sampling of sparse sensor signals},
  author={Bunaiyan, Saleh and Feras Al-Dirini},
  booktitle={2021 IEEE International Midwest Symposium on Circuits and Systems (MWSCAS)},
  pages={1053--1057},
  year={2021},
  organization={IEEE},
  doi={10.1109/MWSCAS47672.2021.9531687}
}

@book{rukhin2001,
  title={A statistical test suite for random and pseudorandom number generators for cryptographic applications},
  author={Rukhin, Andrew and Soto, Juan and Nechvatal, James and Smid, Miles and Barker, Elaine and Leigh, Stefan and Levenson, Mark and Vangel, Mark and Banks, David and Heckert, Alan and others},
  volume={22},
  year={2001},
  publisher={US Dept. of CTA, NIST},
url = {https://tsapps.nist.gov/publication/get_pdf.cfm?pub_id=906762}
}

@article{tokunaga2008true,
  title={True random number generator with a metastability-based quality control},
  author={Tokunaga, Carlos and Blaauw, David and Mudge, Trevor},
  journal={IEEE Journal of Solid-State Circuits
},
  volume={43},
  number={1},
  pages={78--85},
  year={2008},
  publisher={IEEE},
doi={10.1109/JSSC.2007.910965}
}

@inproceedings{brederlow2006low,
  title={A low-power true random number generator using random telegraph noise of single oxide-traps},
  author={Brederlow, Ralf and Prakash, Ramesh and Paulus, Christian and Thewes, Roland},
  booktitle={IEEE Int. Solid-State Circuits Conf. (ISSCC) Dig. Tech. Papers, 2006, pp. 1666--1675 	
},
  pages={1666--1675},
  year={2006},
 doi={10.1109/ISSCC.2006.1696222}
}

@article{akbari2023true,
  title={True random number generator based on the variability of the high resistance state of RRAMs},
  author={Akbari, Maryam and Mirzakuchaki, Sattar and Arum{\'\i}, Daniel and Manich, Salvador and G{\'o}mez-Pau, Alvaro and Campabadal, Francesca and Gonz{\'a}lez, Mireia Bargall{\'o} and Rodr{\'\i}guez-Monta{\~n}{\'e}s, Rosa},
  journal={IEEE Access},
volume={11},
  pages={66682--66693},
  year={2023},
  publisher={IEEE},
doi={10.1109/ACCESS.2023.3290896}
}

@article{piccinini2017self,
  title={Self-heating phase-change memory-array demonstrator for true random number generation},
  author={Piccinini, Enrico and Brunetti, Rossella and Rudan, Massimo},
  journal={IEEE Transactions on Electron Devices},
  volume={64},
  number={5},
  pages={2185--2192},
  year={2017},
  publisher={IEEE},
doi={10.1109/TED.2017.2673867}
}

@article{gong2019true,
  title={True random number generators using electrical noise},
  author={Gong, Lishuang and Zhang, Jianguo and Liu, Haifang and Sang, Luxiao and Wang, Yuncai},
  journal={IEEE Access},
  volume={7},
  pages={125796--125805},
  year={2019},
  publisher={IEEE},
doi={10.1109/ACCESS.2019.2939027}
}

@inproceedings{srinivasan20102,
  title={2.4 GHz 7mW all-digital PVT-variation tolerant true random number generator in 45nm CMOS},
  author={Srinivasan, Suresh and Mathew, Sanu and Ramanarayanan, Rajaraman and Sheikh, Farhana and Anders, Mark and Kaul, Himanshu and Erraguntla, Vasantha and Krishnamurthy, Ram and Taylor, Greg},
  booktitle={Proc. IEEE Symp. VLSI  Circ., 2010, pp.3--4},
  pages={3--4},
  year={2010},
doi={10.1109/VLSIC.2010.5560296}
}

@article{hassan2021quantitative,
  title={Quantitative evaluation of hardware binary stochastic neurons},
  author={Hassan, Orchi and Datta, Supriyo and Camsari, Kerem Y},
  journal={Physical Review Applied},
  volume={15},
  number={6},
  pages={064046},
  year={2021},
  publisher={APS},
doi={10.1103/PhysRevApplied.15.064046}
}

@article{zahoor2024novel,
  title={Novel Postprocessing and Sampling Point Optimization Techniques for Enhancing Quality of Randomness in MTJ-Based TRNGs},
  author={Zahoor, Furqan and Nisar, Arshid and Das, Kunal Kranti and Maitra, Subhamoy and Kaushik, Brajesh Kumar and Chattopadhyay, Anupam},
  journal={IEEE Transactions on Electron Devices},
 volume={71},
  number={7},
  pages={4138-4175},
  year={2024},
  publisher={IEEE},
doi={10.1109/TED.2024.3399170}
}

@article{onizawa2020,
  title={High-throughput/low-energy MTJ-based true random number generator using a multi-voltage/current converter},
  author={Onizawa, Naoya and Mukaida, Shogo and Tamakoshi, Akira and Yamagata, Hitoshi and Fujita, Hiroyuki and Hanyu, Takahiro} ,
  journal={IEEE Transactions on Very Large Scale Integration (VLSI) systems},
  volume={28},
  number={10},
  pages={2171--2181},
  year={2020},
  publisher={IEEE},
doi={10.1109/TVLSI.2020.3005413}
}

@article{frustaci2024high,
  title={A High-Speed and Low-Power DSP-Based TRNG for FPGA Implementations},
  author={Frustaci, Fabio and Spagnolo, Fanny and Corsonello, Pasquale and Perri, Stefania},
  journal={	IEEE Transactions on Circuits and Systems II: Express Briefs
},
volume={17},
  number={12},
  pages={4964--4968},
  year={2024},
  publisher={IEEE},
doi={10.1109/TCSII.2024.3421323}
}

@article{liu2016low,
  title={A low-cost low-power ring oscillator-based truly random number generator for encryption on smart cards},
  author={Liu, Dongsheng and Liu, Zilong and Li, Lun and Zou, Xuecheng},
  journal={	IEEE Transactions on Circuits and Systems II: Express Briefs},
  volume={63},
  number={6},
  pages={608--612},
  year={2016},
  publisher={IEEE},
  doi={10.1109/TCSII.2016.2530800}
}

@article{de2017variation,
  title={A variation-aware timing modeling approach for write operation in hybrid CMOS/STT-MTJ circuits},
  author={De Rose, Raffaele and Lanuzza, Marco and Crupi, Felice and Siracusano, Giulio and Tomasello, Riccardo and Finocchio, Giovanni and Carpentieri, Mario and Alioto, Massimo},
  journal={IEEE Transactions on Circuits and Systems I: Regular Papers},
  volume={65},
  number={3},
  pages={1086--1095},
  year={2017},
  publisher={IEEE},
doi={10.1109/TCSI.2017.2762431}
}

@inproceedings{singh2023hardware,
  title={Hardware security primitives using passive rram crossbar array: Novel trng and puf designs},
  author={Singh, Simranjeet and Zahoor, Furqan and Rajendran, Gokul and Patkar, Sachin and Chattopadhyay, Anupam and Merchant, Farhad},
  booktitle={Proceedings of the 28th Asia and South Pacific Design Automation Conference},
  pages={449--454},
  year={2023},
  doi={10.1145/3566097.3568348}
}

@article{bunaiyan2022neuro,
  title={Neuro-inspired autonomous data acquisition for energy-constrained IoT sensors},
  author={Buniyan, Saleh and Al-Dirini, Feras},
  journal={IEEE Sensors Journal},
  volume={22},
  number={20},
  pages={19466--19479},
  year={2022},
  publisher={IEEE},
doi={10.1109/JSEN.2022.3200627}
}

@inproceedings{raj2024puf,
  title={PUF-based Lightweight Mutual Authentication Protocol for Internet of Things (IoT) Devices},
  author={Raj, Kamal and Bodapati, Srinivasu and Chattopadhyay, Anupam},
  booktitle={2024 IEEE International Symposium on Circuits and Systems (ISCAS)},
  pages={1--5},
  year={2024},
  organization={IEEE},
 doi={10.1109/ISCAS58744.2024.10558672}
}

@article{clemente2024lightweight,
  title={A lightweight chaos-based encryption scheme for IoT healthcare systems},
  author={Clemente-Lopez, Daniel and de Jesus Rangel-Magdaleno, Jose and Mu{\~n}oz-Pacheco, Jesus Manuel},
  journal={Internet of Things},
  volume={25},
  pages={101032},
  year={2024},
  publisher={Elsevier},
  doi={10.1016/j.iot.2023.101032}
}

@ARTICLE{WirelessGeophoneAccess,
  author={Attia, Hussein and Gaya, Sagiru and Alamoudi, Abdullah and M. Alshehri, Fahad and Al-Suhaimi, Abdulrahman and Alsulaim, Nawaf and M. Al Naser, Ahmad and Aghyad Jamal Eddin, Mohamad and M. Alqahtani, Abdullah and Prieto Rojas, Jhonathan and Al-Dharrab, Suhail and Al-Dirini, Feras},
  journal={IEEE Access}, 
  title={Wireless Geophone Sensing System for Real-Time Seismic Data Acquisition}, 
  year={2020},
  volume={8},
  number={},
  pages={81116-81128},
  doi={10.1109/ACCESS.2020.2989280}}

@INPROCEEDINGS{P-Sensing-EDTM,
  author={Bunaiyan, Saleh and Al-Dirini, Feras},
  booktitle={2024 8th IEEE Electron Devices Technology $\&$ Manufacturing Conference (EDTM)}, 
  title={Probabilistic Autonomous Data Acquisition Using Stochastic MTJ Based p-Bits}, 
  year={2024},
  volume={},
  number={},
  pages={1-3},
  doi={10.1109/EDTM58488.2024.10512188}}

@article{PhysRevApplied.17.014016,
  title = {Hardware-Aware In Situ Learning Based on Stochastic Magnetic Tunnel Junctions},
  author = {Kaiser, Jan and Borders, William A. and Camsari, Kerem Y. and Fukami, Shunsuke and Ohno, Hideo and Datta, Supriyo},
  journal = {Phys. Rev. Appl.},
  volume = {17},
  issue = {1},
  pages = {014016},
  numpages = {12},
  year = {2022},
  month = {Jan},
  publisher = {American Physical Society},
  doi = {10.1103/PhysRevApplied.17.014016},
  url = {https://link.aps.org/doi/10.1103/PhysRevApplied.17.014016}
}

@article{SubnanosecondFluctuations,
  title = {Subnanosecond Fluctuations in Low-Barrier Nanomagnets},
  author = {Kaiser, J. and Rustagi, A. and Camsari, K. Y. and Sun, J. Z. and Datta, S. and Upadhyaya, P.},
  journal = {Phys. Rev. Appl.},
  volume = {12},
  issue = {5},
  pages = {054056},
  numpages = {9},
  year = {2019},
  month = {Nov},
  publisher = {American Physical Society},
  doi = {10.1103/PhysRevApplied.12.054056},
  url = {https://link.aps.org/doi/10.1103/PhysRevApplied.12.054056}
}

@article{NanosecondRTN,
  title = {Nanosecond Random Telegraph Noise in In-Plane Magnetic Tunnel Junctions},
  author = {Hayakawa, K. and Kanai, S. and Funatsu, T. and Igarashi, J. and Jinnai, B. and Borders, W. A. and Ohno, H. and Fukami, S.},
  journal = {Phys. Rev. Lett.},
  volume = {126},
  issue = {11},
  pages = {117202},
  numpages = {6},
  year = {2021},
  month = {Mar},
  publisher = {American Physical Society},
  doi = {10.1103/PhysRevLett.126.117202},
  url = {https://link.aps.org/doi/10.1103/PhysRevLett.126.117202}
}

@article{baturone2022unified,
  title={A unified multibit PUF and TRNG based on ring oscillators for secure IoT devices},
  author={Baturone, Iluminada and Rom{\'a}n, Roberto and Corbacho, {\'A}ngel},
  journal={IEEE Internet of Things Journal},
  volume={10},
  number={7},
  pages={6182--6192},
  year={2022},
  publisher={IEEE},
doi={10.1109/JIOT.2022.3224298}
}

@article{wallace2016toward,
  title={Toward sensor-based random number generation for mobile and IoT devices},
  author={Wallace, Kyle and Moran, Kevin and Novak, Ed and Zhou, Gang and Sun, Kun},
  journal={IEEE Internet of Things Journal},
  volume={3},
  number={6},
  pages={1189--1201},
  year={2016},
  publisher={IEEE},
doi={10.1109/JIOT.2016.2572638}
}

@inproceedings{rajendran2024harnessing,
  title={Harnessing Entropy: RRAM Crossbar-based Unified PUF and RNG},
  author={Rajendran, Gokulnath and Zahoor, Furqan and Thakker, Sidhaant Sachin and Singh, Simranjeet and Merchant, Farhad and Rana, Vikas and Chattopadhyay, Anupam},
  booktitle={2024 37th International Conference on VLSI Design and 2024 23rd International Conference on Embedded Systems (VLSID)},
  pages={560--564},
  year={2024},
  organization={IEEE},
doi={10.1109/VLSID60093.2024.00099}
}

@ARTICLE{Mohammed_2025,
  author={Mohammed, Mahmood A. and Al-Dirini, Feras and Emara, Ahmed S. and Roberts, Gordon W.},
  journal={IEEE Transactions on Circuits and Systems I: Regular Papers}, 
  title={Design for Slew-Rate in Multi-Stage CMOS OTAs}, 
  year={2026},
  volume={73},
  number={2},
  pages={775-788},
  doi={10.1109/TCSI.2025.3593405}}

@misc{PatentDecoupledPBit,
  author       = {Al-Dirini, Feras Mohamad Ameer and Bunaiyan, Saleh Ahmad S.},
  title        = {P-bit generator and methods for tuning a P-bit generator having decoupled stochastic and control paths},
  howpublished = {U.S.\ Patent Appl.\ No.\ 18\,949\,434},
  year         = {2025},
  note         = {Filed 15 Nov 2024},
  url          = {https://patents.google.com/patent/US20250167774A1/en}
}

@misc{PatentTunablePBit,
  author       = {Bunaiyan, Saleh Ahmad S. and Al-Dirini, Feras Mohamad Ameer},
  title        = {Apparatus and method of implementing a probabilistic bit (P-bit) circuit with enhanced tunability},
  howpublished = {U.S.\ Patent Appl.\ No.\ 18\,417\,631},
  year         = {2025},
  note         = {Filed 19 Jan 2024},
  url          = {https://patents.google.com/patent/US20250169370A1/en}
}

@article{MemristorPUFFScienceAdv,
author = {Xueqi Li  and Bohan Lin  and Bin Gao  and Yuyao Lu  and Siyao Yang  and Zhiqiang Su  and Ting-Ying Shen  and Jianshi Tang  and He Qian  and Huaqiang Wu },
title = {A memristor-based unified PUF and TRNG chip with a concealable ability for advanced edge security},
journal = {Science Advances},
volume = {11},
number = {13},
pages = {eadr0112},
year = {2025},
doi = {10.1126/sciadv.adr0112},
URL = {https://www.science.org/doi/abs/10.1126/sciadv.adr0112},
}

@misc{P-NeuronsISCAS,
      title={Configurable p-Neurons Using Modular p-Bits}, 
      author={Saleh Bunaiyan and Mohammad Alsharif and Abdelrahman S. Abdelrahman and Hesham ElSawy and Suraj S. Cheema and Suhaib A. Fahmy and Kerem Y. Camsari and Feras Al-Dirini},
      year={2026},
      eprint={2601.18943},
      archivePrefix={arXiv},
      primaryClass={cs.ET},
      url={https://arxiv.org/abs/2601.18943}, 
}

@misc{P-SensingISCAS,
      title={Probabilistic Sensing: Intelligence in Data Sampling}, 
      author={Ibrahim Albulushi and Saleh Bunaiyan and Suraj S. Cheema and Hesham ElSawy and Feras Al-Dirini},
      year={2026},
      eprint={2601.19953},
      archivePrefix={arXiv},
      primaryClass={cs.LG},
      url={https://arxiv.org/abs/2601.19953}, 
}

@misc{PatentPSensing,
  author       = {Al-Dirini, Feras Mohamad Ameer and Bunaiyan, Saleh Ahmad S.},
  title        = {Probabilistic autonomous data acquisition using stochastic mtj based p-bits},
  howpublished = {U.S.\ Patent Appl.\ No.\ 18\,417\,677},
  year         = {2025},
  note         = {Filed 19 Jan 2024},
  url          = {https://patents.google.com/patent/US20250238290A1/en}
}

@misc{PatentPNeurons,
  author       = {Al-Dirini, Feras Mohamad Ameer and Bunaiyan, Saleh Ahmad S. and Alsharif, Mohammad Sharaf F.},
  title        = {Modular Probabilistic Bit for Implementing Stochastic Neurons with Configurable Activation Functions},
  howpublished = {U.S.\ Patent Appl.\ No.\ 19\,260\,798},
  year         = {2026},
  note         = {Filed 07 Jul 2025},
  url          = {}
}

@misc{PatentAdaptiveTRNG,
  author       = {Al-Dirini, Feras Mohamad Ameer and Zahoor, Furqan and Albulushi, Ibrahim Abdulrahman},
  title        = {Adaptive Random Number Generator for Cryptographic Application},
  howpublished = {U.S.\ Patent Appl.\ No.\ 19\,170\,745},
  year         = {2026},
  note         = {Filed 04 Apr 2025},
  url          = {}
}

@INPROCEEDINGS{SNWMemristor,
  author={Abdelrahman, Abdelrahman S. and ElSawy, Hesham and Lanza, Mario and Akinwande, Deji and Al-Dirini, Feras},
  booktitle={2023 Silicon Nanoelectronics Workshop (SNW)}, 
  title={Scalability of h-BN Based Memristors: Yield and Variability Considerations}, 
  year={2023},
  volume={},
  number={},
  pages={109-110},
  doi={10.23919/SNW57900.2023.10183973}}

@misc{RSQMemristorScaling,
  author = {Al-Dirini, Feras and Abdelrahman, Abdelrahman and ElSawy, Hesham and Yuan, Yue and Cheema, Suraj and Akinwande, Deji and Lanza, Mario},
  title  = {Defect-Aware Extreme Device Scaling Limits of 2D Memristive Technologies},
  year   = {2025},
  note   = {04 August 2025, Preprint (Version 1) available at Research Square [\url{https://doi.org/10.21203/rs.3.rs-7265912/v1}]},
  doi    = {10.21203/rs.3.rs-7265912/v1},
  url    = {https://doi.org/10.21203/rs.3.rs-7265912/v1}
}

@article{cao2021new,
  title={A new energy-efficient and high throughput two-phase multi-bit per cycle ring oscillator-based true random number generator},
  author={Cao, Yuan and Zhao, Xiaojin and Zheng, Wenhan and Zheng, Yue and Chang, Chip-Hong},
  journal={IEEE Transactions on Circuits and Systems I: Regular Papers},
  volume={69},
  number={1},
  pages={272--283},
  year={2021},
  publisher={IEEE},

doi={10.1109/TCSI.2021.3087512}
}

@article{luo20232,
  title={A 2.5 pJ/bit PVT-tolerant true random number generator based on native-NMOS-regulated ring oscillator},
  author={Luo, Yian and Zhang, Junhang and Hao, Jiacheng and Zhao, Xiaojin},
  journal={IEEE Transactions on Circuits and Systems II: Express Briefs},
  volume={70},
  number={10},
  pages={3927--3931},
  year={2023},
  publisher={IEEE},

doi={10.1109/TCSII.2023.3288036}
}

@article{bae20163,
  title={3-Gb/s high-speed true random number generator using common-mode operating comparator and sampling uncertainty of D flip-flop},
  author={Bae, Sang-Geun and Kim, Yongtae and Park, Yunsoo and Kim, Chulwoo},
  journal={IEEE Journal of Solid-State Circuits},
  volume={52},
  number={2},
  pages={605--610},
  year={2016},
  publisher={IEEE},

doi={10.1109/JSSC.2016.2625341}
}

@article{kim2019nano,
  title={Nano-intrinsic true random number generation: A device to data study},
  author={Kim, Jeeson and Nili, Hussein and Truong, Nhan Duy and Ahmed, Taimur and Yang, Jiawei and Jeong, Doo Seok and Sriram, Sharath and Ranasinghe, Damith C and Ippolito, Samuel and Chun, Hosung and others},
  journal={IEEE Transactions on Circuits and Systems I: Regular Papers},
  volume={66},
  number={7},
  pages={2615--2626},
  year={2019},
  publisher={IEEE},

doi={10.1109/TCSI.2019.2895045}
}

@article{arumi2023true,
  title={True random number generator based on rram-bias current starved ring oscillator},
  author={Arumi, Daniel and Manich, S and G{\'o}mez-Pau, A and Rodriguez-Montanes, Rosa and Gonz{\'a}lez, MB and Campabadal, Francesca},
  journal={IEEE Journal on Exploratory Solid-State Computational Devices and Circuits},
  volume={9},
  number={2},
  pages={92--98},
  year={2023},
  publisher={IEEE},

doi={10.1109/JXCDC.2023.3320056}
}

@article{fu2023rhs,
  title={RHS-TRNG: A resilient high-speed true random number generator based on STT-MTJ device},
  author={Fu, Siqing and Li, Tiejun and Zhang, Chunyuan and Li, Hanqing and Ma, Sheng and Zhang, Jianmin and Zhang, Ruiyi and Wu, Lizhou},
  journal={IEEE Transactions on Very Large Scale Integration (VLSI) Systems},
  volume={31},
  number={10},
  pages={1578--1591},
  year={2023},
  publisher={IEEE},

doi={10.1109/TVLSI.2023.3298327}
}

@article{koh2025closed,
  title={Closed Loop Superparamagnetic Tunnel Junctions for Reliable True Randomness and Generative Artificial Intelligence},
  author={Koh, Dooyong and Wang, Qiuyuan and McGoldrick, Brooke C and Chou, Chung-Tao and Liu, Luqiao and Baldo, Marc A},
  journal={Nano Letters},
  volume={25},
  number={10},
  pages={3799--3806},
  year={2025},
  publisher={ACS Publications},

doi={10.1021/acs.nanolett.4c05728}
}

@article{marsaglia2002some,
  title={An asynchronous and low-power true random number generator using STT-MTJ},
  author={Marsaglia, George and Tsang, Wai Wan },
  journal={Journal of Statistical Software},
  volume={7},
  number={},
  pages={1--9},
  year={2002},
  publisher={},

doi={10.18637/jss.v007.i03}
}

@article{l2007testu01,
  title={TestU01: AC library for empirical testing of random number generators},
  author={ L'ecuyer, Pierre and Simard, Richard },
  journal={ACM Transactions on Mathematical Software (TOMS)},
  volume={33},
  number={4},
  pages={1--40},
  year={2007},
  publisher={ACM New York, NY, USA},

doi={10.1145/1268776.1268777}
}

@article{foreman2024statistical,
  title={Statistical testing of random number generators and their improvement using randomness extraction},
  author={ Foreman, Cameron and Yeung, Richie and Curchod, Florian J},
  journal={Entropy},
  volume={26},
  number={12},
  pages={1053},
  year={2024},
  publisher={MDPI},

doi={10.3390/e26121053}
}

@article{foreman2023practical,
  title={Practical randomness amplification and privatisation with implementations on quantum computers},
  author={ Foreman, Cameron and Wright, Sherilyn and Edgington, Alec and Berta, Mario and Curchod, Florian J},
   journal={Quantum},
  volume={7},
  pages={969},
  year={2023},
  publisher={Verein zur F{\"o}rderung des Open Access Publizierens in den Quantenwissenschaften},

doi={10.22331/q-2023-03-30-969}
}

@article{sleem2020testu01,
  title={TestU01 and Practrand: Tools for a randomness evaluation for famous multimedia ciphers},
  author={Sleem, Lama and Couturier, Rapha{\"e}l},
   journal={Multimedia Tools and Applications},
  volume={79},
  number={33},
  pages={24075--24088},
  year={2020},
  publisher={Springer},

doi={10.1007/s11042-020-09108-w}
}

\end{document}